\documentclass[12pt, letterpaper]{article}

\usepackage{amssymb}
\usepackage{amsmath}
\usepackage{verbatim}

\usepackage[nosort]{cite}
\usepackage[hyperref,bulletsep]{collect}
\usepackage{hyperref}

\mathversion{normal}

\makeatletter \@addtoreset{equation}{section} \makeatother

\newcommand{\gen}[1]{\mathfrak{#1}}
\newcommand{\alg}[1]{\mathfrak{#1}}

\ifx\genfrac\sdflkaj

\else

\fi
\newcommand{\sfrac}[2]{{\textstyle\frac{#1}{#2}}}
\ifx\half\sdflkaj
\newcommand{\half}{\sfrac{1}{2}}
\fi

\newcommand{\comm}[2]{[#1,#2]}

\let\oldPhi=\Phi
\let\oldPsi=\Psi
\renewcommand{\Phi}{\mathnormal{\oldPhi}}
\renewcommand{\Psi}{\mathnormal{\oldPsi}}

\newcommand{\nln}{\nonumber\\}
\newcommand{\nl}{\nonumber\\&&\mathord{}}
\newcommand{\earel}[1]{\mathrel{}&#1&\mathrel{}}
\newcommand{\eq}{\earel{=}}
\newenvironment{myeqnarray}{\arraycolsep0pt\begin{eqnarray}}{\end{eqnarray}\ignorespacesafterend}
\newenvironment{myeqnarray*}{\arraycolsep0pt\begin{eqnarray*}}{\end{eqnarray*}\ignorespacesafterend}

\def\[{\begin{equation}}
\def\]{\end{equation}}
\def\<{\begin{myeqnarray}}
\def\>{\end{myeqnarray}}

\ifx\href\asklfhas\newcommand{\href}[2]{#2}\fi
\newcommand{\arxivno}[1]{\href{http://arxiv.org/abs/#1}{#1}}

\begin{document}
\thispagestyle{empty}
\begin{flushright}\footnotesize
\texttt{\arxivno{arXiv:1111.1773}}\\ 
\texttt{CALT-68-2853}
\end{flushright}
\vspace{.5cm}

\begin{center}%
{\Large\textbf{\mathversion{bold}%
$C$harging the Superconformal Index  }\par}\vspace{1.5cm}%

\textsc{Benjamin~I.~Zwiebel} \vspace{8mm}

\textit{ California Institute of Technology \\ 
Pasadena, CA 91125, USA}%
\vspace{4mm}

\texttt{bzwiebel@caltech.edu}\par\vspace{1.5cm}

\textbf{Abstract}\vspace{7mm}

\begin{minipage}{14.7cm}

The superconformal index is an important invariant of superconformal field theories.  In this note we refine the superconformal index by inserting the charge conjugation operator $C$. We construct a matrix integral for this charged index for $\mathcal{N}=4$ SYM with $SU(N)$ gauge group. The key ingredient for the construction is  a ``charged character,''  which reduces to $\mathrm{Tr} (C)$ for singlet representations of the gauge group. For each irreducible real $SU(N)$ representation, we conjecture that this charged character is equal to the standard character for a corresponding representation of $SO(N+1)$ or $SP(N-1)$, for $N$ even or odd respectively.  The matrix integral for the charged index passes tests for small $N$ and for $N \rightarrow \infty$. Like the ordinary superconformal index, for $\mathcal{N}=4$ SYM the charged index is independent of $N$ in the large-$N$ limit.
\end{minipage}

\end{center}

\newpage
\setcounter{page}{1}
\renewcommand{\thefootnote}{\arabic{footnote}}
\setcounter{footnote}{0}


\section{Introduction}

In recent years, the superconformal index \cite{Kinney:2005ej} has been a central object for investigations of dualities for superconformal field theories. The superconformal index gives a weighted counting of BPS states, and this topological invariant is equal for theories related by dualities. In four dimensions, this index has been used to study the $AdS$/CFT correspondence, Seiberg duality, and S-dualities.  \cite{Kinney:2005ej} confirmed the matching of the superconformal indices  for large-$N$ $\mathcal{N}=4$ SYM and the dual supergravity in $AdS_5 \times S^5$, and a similar agreement exists for supergravity theory in $AdS_5 \times T^{1,1}$ \cite{Nakayama:2006ur} and the  dual conifold gauge theory \cite{Gadde:2010en}. The computation of the index for $\mathcal{N}=1$ superconformal field theories arising as IR fixed points \cite{Romelsberger:2005eg,Romelsberger:2007ec}, recently justified rigorously \cite{Festuccia:2011ws}, has enabled many tests of Seiberg duality \cite{Dolan:2008qi,Spiridonov:2008zr} and yielded rich connections to the theory of elliptic hypergeometric integrals \cite{Spiridonov:2001aa,Spiridonov:2008aa}, including new conjectured identities\cite{Spiridonov:2009za}.   The superconformal index also has been studied for a large class of non-Lagrangian $\mathcal{N}=2$ superconformal  gauge theories introduced in  \cite{Gaiotto:2009we}. In this case, due to S-duality,  the index can be computed using a two-dimensional topological QFT \cite{Gadde:2009kb}, which has been identified in various limits \cite{Gadde:2011uv}.  

The importance of the superconformal index motivates us to consider refinements to it that will enable further probes of superconformal theories. While the superconformal index of \cite{Kinney:2005ej} includes all information about protected short representations that follows from superconformal symmetry alone, when additional symmetries are present further refinements of the index may be possible. For example, for theories with charge conjugation symmetry,  we  can modify the index by inserting the charge conjugation operator $C$, and we call this refined index the charged (superconformal) index or $\mathcal{I}_C$. Such a refinement of the closely-related Witten index was first considered many years ago \cite{Witten:1982df}. Because $C$ acts on the gauge group representations, the matrix integral for the superconformal index  \cite{Kinney:2005ej} does not apply to $\mathcal{I}_C$, and  this note focuses on finding and testing a matrix integral for the charged superconformal index. We use $\mathcal{N}=4$ SYM with $SU(N)$ gauge group as our example, but we do not expect obstacles to generalizing our approach to other superconformal gauge theories (with $C$ symmetry and with a weak-coupling limit or dual). 

The integrand of the matrix integral for the superconformal index expands in products of gauge group characters, where the characters correspond to the gauge group representations of the fundamental letters of the theory. Integration over the gauge group then  projects onto gauge group singlet states, due to the orthonormality of characters.  We use a similar approach for computing $\mathcal{I}_C$ for $SU(N)$ gauge group. This requires the introduction of a ``charged character'' that is the trace of an operator $K_C(z_1, z_2, \ldots z_{[N/2]})$. $K_C$ is an exponential multiplied by $C$. The argument of the exponential is a weighted sum of mutually commuting raising operators for the $\alg{su}(N)$ algebra, and the $z_i$ appear in the coefficients of the raising operators. Importantly, $K_C$ then reduces to $C$ for singlet representations. The key identity that we conjecture relates the charged character to ordinary characters of orthogonal or symplectic groups as
\[  \label{eq:keyidentity}
\mathrm{Tr}_{\mathbf{d}} K_C(z_1,z_2 \ldots z_{[N/2]}) =  \chi^{G(N)}_{p(\mathbf{d})}(z_1, z_2 \ldots z_{[N/2]}), 
\]
where $G(N)$ is $SO(N+1)$ for even $N$ and $SP(N-1)$ for odd $N$. 
The left side is the trace of $K_C$ over a real $SU(N)$ irreducible representation with the Dynkin labels $\mathbf{d}$.  On the right side, we have the ordinary character for the $G(N)$ irreducible representation that has Dynkin labels $p (\mathbf{d})$, where $p$ is a simple one-to-one map between Dynkin labels for real irreducible representations of $SU(N)$ and for irreducible representations of $G(N)$.  Because of the relation (\ref{eq:keyidentity}) to ordinary characters, the charged characters for different irreducible representations are orthonormal with respect to the $G(N)$ Haar measure. Therefore, in analogy to the ordinary superconformal index, it is straightforward to construct the matrix integral that gives $\mathcal{I}_C$ for adjoint letters using this charged character in place of ordinary characters and integrating over $G(N)$ rather than $SU(N)$. 

The remainder of this note is organized as follows. Section \ref{sec:chargedindex} reviews the character derivation of the matrix integral for the ordinary superconformal index, and then gives the analogous construction for $\mathcal{I}_C$ using the charged character. We describe tests that the construction passes for small and large $N$ in Sections \ref{sec:smallN} and \ref{sec:largeN} respectively. Some open problems are discussed in Section \ref{sec:concl}. Appendix \ref{sec:appformulas} includes useful formulas for characters and measures, and Appendix \ref{sec:appsu2su3}  and Appendix \ref{sec:appdata} give data for some tests at small $N$.
\section{Calculating the charged index \label{sec:chargedindex}}

 This section starts with a review of the superconformal index and of its computation using characters. We give many basic details of this character calculation to prepare for the second part of the section, where we introduce a charged character that can be used to compute the charged superconformal index in a similar way.

\subsection{Review of the character calculation of the superconformal index}

The superconformal index \cite{Kinney:2005ej} is defined for superconformal field theories on $S^3 \times R$ as 
\[
\mathrm{Tr} \big( \, (-1)^F e^{\mu_i q_i}\big), \label{eq:superconformalindex}
\]
where $F$ is the fermion number, and the $\mu_i$ are chemical potentials for the conserved charges $q_i$ that commute with a Hermitian conjugate pair of supercharges. The trace is over all physical (gauge-invariant) states of this radially quantized theory.

For $\mathcal{N}=4$ SYM with gauge group $SU(N)$, using the notation of \cite{Kinney:2005ej}, (\ref{eq:superconformalindex}) becomes
\[
\mathrm{Tr} \big( \,  (-1)^F e^{- \beta \Delta}t^{2(E + j_1)} y^{2 j_2} v^{R_2} w^{R_3} \big). \label{eq:n=4index}
\]
Here $\Delta$ is twice the anticommutator of the above-mentioned conjugate pair of supercharges, $E$ is the energy, $j_1$ and $j_2$ are  the representation labels for the $SO(4)$ symmetry of $S^3$, and $R_2$ and $R_3$ are Cartan charges of the $SU(4)$ $R$-symmetry group. $t$, $y$, $v$, $w$ are exponentials of the chemical potentials. The index is independent of $\beta$ since the contributions from states with $\Delta \neq 0$ cancel in boson-fermion pairs. The trace over physical states requires us to count singlet representations of $SU(N)$. 

Since the index is invariant under continuous deformations that preserve superconformal invariance, we can evaluate the index using free $\mathcal{N}=4$ SYM with `t Hooft coupling $\lambda=0$. Then the states  are simply $SU(N)$ singlet (linear) combinations of the single-excitation ``letters,'' where these letters correspond to the excitation modes on $S^3$ (decoupled harmonic oscillators for the free theory). These letters transform in the adjoint representation of $SU(N)$. As we will review below, after doing this counting, we find that the index depends on the letters only through the $(\Delta=0)$ single-letter index f, 
\[
f= \sum_{\text{letters}}   (-1)^F t^{2(E + j_1)} y^{2 j_2} v^{R_2} w^{R_3}. \label{eq:definef}
\]
\cite{Kinney:2005ej} evaluated $f$, obtaining
\[
f(t, y, v, w) = \frac{t^2 (v + \frac{1}{w} + \frac{w}{v}) - t^3 (y + \frac{1}{y}) - t^4 (w + \frac{1}{v}+\frac{v}{w}) + 2 t^6}{(1-t^3 y)(1 - \frac{t^3}{y})} \label{eq:n=4f}
\]
and 
\[ \label{eq:1minusf}
1-f = \frac{(1 - t^2/w)(1 - t^2 w / v)(1 - t^2 v)}{(1-t^3 y)(1 - t^3 /y)},
\]
which is always positive for the allowed values of chemical potentials\footnote{Convergence of the index requires that every letter has weight of absolute value less than one, and this requirement for the scalars and derivatives implies that all factors for $1-f$ are positive.}.

In preparation for counting gauge-invariant states, we will review characters for $SU(N)$. First, we parameterize the Lie algebra  $\alg{su}(N)$ by  $N^2$ $\gen{J}^A_B$ ($A,B=1,2 \ldots N$) that satisfy
\[
\comm{\gen{J}^{A}_B}{\gen{J}^{C}_D}=  \delta^A_D \, \gen{J}^C_B - \delta^C_B \, \gen{J}^A_D, \quad \sum_{A=1}^N \gen{J}^A_A = 0, \quad \gen{J}^A_B = (\gen{J}^B_A)^\dagger.
\]
We also choose a basis for the $N-1$ Cartan generators $\gen{J}_i$ as
\[
\gen{J}_i = \gen{J}^i_i - \gen{J}^{i+1}_{i+1}, \quad i = 1, 2 \ldots N-1,
\]
where we do not sum over $i$ on the right side. The $\gen{J}^A_B$ for $A \neq B$ are the raising and lowering operators; the remaining $(N^2-N)$ non-Cartan  Hermitian generators are proportional to the symmetric and antisymmetric combinations $(\gen{J}^A_B + \gen{J}^B_A)$ and $ i (\gen{J}^A_B - \gen{J}^B_A)$ for $A \neq B$.  We label irreducible representations of $SU(N)$ as $R_\mathbf{d}$, where  the Dynkin labels $\mathbf{d}$ are given by
\[
\mathbf{d} = [d_1, d_2, \ldots d_{N-1}], \quad d_i = J_i,
\]
and $J_i$ is the Cartan generator $\gen{J}_i$ eigenvalue of the highest-weight element of the representation. 

We parameterize an element of the Cartan subgroup  using  $z_i$, $i=1, 2 \ldots (N-1)$, 
\[ \label{eq:defineK}
K(z_i) = \mathrm{exp} \Big(\sum_{j=1}^{N-1} i \theta_j (\gen{J}_j^j-\gen{J}_N^N) \Big), \quad e^{i \theta_j} = z_j.
\]
Then the character $\chi_{\mathbf{d}}$ of an irreducible representation $R_\mathbf{d}$  is simply the trace over the representation of $K(z_i)$
\[
\chi_\mathbf{d}(z_i) = \mathrm{Tr}_{R_\mathbf{d}} K(z_i). \label{eq:definecharacter}
\]
The Weyl character formula leads to a simple determinant formula for $\chi_\mathbf{d}(z_i)$, (\ref{eq:suncharacter})\footnote{As explained in Appendix \ref{sec:appformulas}, it will turn out to be useful to instead write $SU(N)$ characters as homogeneous functions of $N$ $z_i$. Setting $z_N = (z_1 z_2 \ldots z_{N-1})^{-1}$ in such expressions yields the form described here.}. For our counting problem, $\chi_\mathbf{d}$ has three key properties. The first two are its additive and multiplicative properties, which follow immediately from its definition as a trace:
\[
\chi_{\mathbf{d_1} \oplus \mathbf{d_2} } = \chi_\mathbf{d_1}  +  \chi_\mathbf{d_2}, \quad \chi_{\mathbf{d_1} \otimes \mathbf{d_2} } = \chi_\mathbf{d_1}  \cdot  \chi_\mathbf{d_2}. \label{eq:characterproperties}
\]
 Here $\chi_{\mathbf{d_1} \oplus \mathbf{d_2} } $ is the character for the representation $R_{\mathbf{d_1}} \oplus R_{\mathbf{d_2}}$ and similarly for $\chi_{\mathbf{d_1} \otimes \mathbf{d_2} }$. The third property is the orthonormality of the characters with respect to the Haar measure for $SU(N)$. The Haar measure $\mathrm{d}\mu(z_i)$ is given  in (\ref{eq:sunmeasure}). In particular, we have
 \[
 \int_{SU(N)} \mathrm{d}\mu(z_i) \chi_\mathbf{d}(z_i)= \delta_{\mathbf{0},\mathbf{d}},
 \]
 where the right side is one when $R_\mathbf{d}$ is the singlet representation and 0 otherwise.

Working toward the complete evaluating of the index for $\mathcal{N}=4$ SYM (\ref{eq:n=4index}), let us first consider a model with a single bosonic letter transforming in the adjoint, and with weight $x_B$. The index (or partition function) $\mathcal{I}$  for such a model is simply \cite{Aharony:2003sx}
\[
\mathcal{I}(x_B) = \int _{SU(N)} \mathrm{d}\mu(z_i) \frac{1}{\mathrm{det}_{\text{adjoint}} \big(1 - x_B \, K(z_i)\big)}. \label{eq:singlebosonindex}
\]
The integrand  accounts for repeated letters appearing in totally symmetric representations. A short derivation can be found in Appendix A of \cite{Aharony:2003sx}. For later application, it is important to note that this determinant formula could be used similarly for any operator $K'$ that can be represented by a $(N^2-1) \times (N^2-1)$ complex matrix, and does not require that $K'$ be a group element.  The integration in (\ref{eq:singlebosonindex}) projects to the physical singlet states. 

We can rewrite the integrand in a useful way. First label the $(N^2-1)$  eigenvalues of the adjoint representation of $K$ as $k_i$. Then the integrand becomes
\[
\frac{1}{\mathrm{det}_{\text{adjoint}} \big(1 - x_B \, K(z_i)\big)} = \prod_{i=1}^{N^2-1} \frac{1}{1 - x_B k_i}. \label{eq:detineigenbasis}
\]
We can further simplify the the right side as 
\<
 \prod_{i=1}^{N^2-1} \frac{1}{1 - x_B k_i}  \eq  \prod_{i=1}^{N^2-1}  \mathrm{exp}  \big(- \mathrm{log}(1 - x_B k_i)  \big) =  \prod_{i=1}^{N^2-1}  \mathrm{exp} \Big( \sum_{m=1}^\infty \frac{(x_B k_i)^m}{m} \Big)  \nln
\eq  \mathrm{exp} \Big( \sum_{m=1}^\infty \frac{x_B^m}{m} \big(\sum_{i=1}^{N^2-1} k_i^m \big) \Big)  = \mathrm{exp} \Big( \sum_{m=1}^\infty \frac{x_B^m}{m} \chi_{\text{adjoint}}(z_i^m) \Big). 
\>
The last equality follows since the eigenvalues $k_i$ are monomials in the $z_i$, and since the sum of the $k_i$ is simply  $\mathrm{Tr} K = \chi$. Then we have
\<
\mathcal{I}(x_B) \eq  \int_{SU(N)} \mathrm{d}\mu(z_i) \frac{1}{\mathrm{det}_{\text{adjoint}} \big(1 - x_B \, K(z_i)\big)}
\nln
\eq   \int_{SU(N)} \mathrm{d}\mu(z_i) \, \mathrm{exp} \Big( \sum_{m=1}^\infty \frac{x_B^m}{m} \chi_{\text{adjoint}}(z_i^m) \Big). \label{eq:bosonicexpform}
\>

Similarly, for a single fermionic letter with weight $x_F$ we have
\<
\mathcal{I}(x_F) \eq  \int_{SU(N)} \mathrm{d}\mu(z_i) \, \mathrm{det}_{\text{adjoint}} \big(1 - x_F \, K(z_i)\big)
\nln
\eq   \int_{SU(N)} \mathrm{d}\mu(z_i)\,  \mathrm{exp} \Big( \sum_{m=1}^\infty \frac{-x_F^m}{m} \chi_{\text{adjoint}}(z_i^m) \Big), \label{eq:fermionicexpform}
\>
where the integrand corresponds to the antisymmetric products of adjoint representations appropriate for fermions, and also includes the appropriate sign for the $(-1)^F$ factor.

Using the product property of characters of (\ref{eq:characterproperties}), it follows that for each bosonic (fermionic) letter of $\mathcal{N}=4$ SYM with weight $x_B$ ($x_f$)  we have a factor of the same form as the integrand  of (\ref{eq:bosonicexpform})  ((\ref{eq:fermionicexpform})). The full index is then given by the $SU(N)$ integration of the product of these factors for all $(\Delta=0)$ letters of $\mathcal{N}=4$ SYM. This simplifies to depend only on $f$ (\ref{eq:definef} - \ref{eq:n=4f}) as 
\[
\mathcal{I}(t, y,v, w) =   \int_{SU(N)} \mathrm{d}\mu(z_i)  \, \mathrm{exp} \Big( \sum_{m=1}^\infty \frac{f(t^m, y^m, v^m, w^m)}{m} \chi_{\text{adjoint}}(z_i^m) \Big).
\]
%

\subsection{Computing the charged index using a charged character \label{sec:computingchargedindex}}

For a superconformal  field theory with $C$ symmetry, we define the  the charged superconformal index $\mathcal{I}_C$ as   
\[
\mathcal{I}_C = \mathrm{Tr}\big(\, (-1)^F C \, e^{\mu_i q_i}\big).
\]
The $q_i$ must commute with the charge conjugation operator $C$. Since $C$ commutes with the supersymmetry generators, the same arguments as for the ordinary index imply that $\mathcal{I}_C$ is in fact an index. For  $\mathcal{N}=4$ SYM with $SU(N)$ gauge group,  charge conjugation acts simply  as complex conjugation of $SU(N)$ representations.  So $C$ commutes with all global symmetries, and  $\mathcal{I}_C$ can include all of the same chemical potentials as the superconformal index. In terms of the $\gen{J}^A_B$ parameterizing $\alg{su}(N)$ we have 
\[
C \, \gen{J}^A_B \, C = -\gen{J}^B_A.
\]
The letters of  $\mathcal{N}=4$ SYM transform in the adjoint representation. Acting on a letter corresponding to a particular $\alg{su}(N)$ generator, $C$ returns minus the letter corresponding to the Hermitian conjugate generator, which equals minus the transpose.  For example, the adjoint representation of $C$ for $SU(2)$ and $SU(3)$ is given in Appendix \ref{sec:appsu2su3}. It follows that we can no longer use $SU(N)$ characters to compute the trace over physical states. Nonetheless, we can compute $\mathcal{I}_C$ in an analogous way, once we define a charged character for $SU(N)$. 

In analogy to the group element $K(z_i)$ (\ref{eq:defineK}) whose trace gives the $SU(N)$ character, we define an operator 
\[
K_{C} (z_1, z_2 \ldots z_{[N/2]}) = \exp \Big(\sum_{i=1}^{[N/2]} (z_i^{1/2} + z_i^{-1/2} ) \gen{J}^{2 i-1}_{2 i}\Big)C. \label{eq:definekc}
\]
$[N/2]$ is the greatest integer less than or equal to $N/2$. Since we will only use the trace of $K_C$ over representations, we have lots of freedom to pick the $[N/2]$ raising or lowering operators  and obtain the same results below. Here we choose $\gen{J}^{2 i-1}_{2 i}$, but we could use any set of $[N/2]$ raising or lowering operators such that each operator and its Hermitian conjugate commute with all the other operators and their conjugates.  Also, note that for any (necessarily finite-dimensional) representation the exponential expands as a finite sum since it is built from raising or lowering operators (this is why $z_i$ rather than $\theta_i$ appears in the exponential).  $K_C$ is clearly not a group element, but it still can be represented by a  $d\times d$ complex matrix for any $d$-dimensional representation of $SU(N)$.

We conjecture that $K_C$ has the following essential property. For irreducible real representations, the trace of $K_C$ is equal to the character of an irreducible representation of an orthogonal or symplectic group. Therefore, we call $\mathrm{Tr}(K_C)$ the charged character. To state this relation precisely, we introduce a projector that maps $(N-1)$-dimensional Dynkin labels to $[N/2]$-dimensional ones as
\[ \label{eq:definep}
p(\mathbf{d}) = [d_1, d_2 \ldots d_{[N/2]}].
\]
 Note that $\mathrm{Tr}(K_C)$ vanishes for a representation that is a direct sum of a complex representation and its complex conjugate, since $C$ then maps elements from one representation to the complex conjugate one. According to our conjecture, the charged character for the real $SU(N)$ representation $R_\mathbf{d}$ is\footnote{When considering possible representations of $SU(N)$ in the abstract, we are free to choose an overall sign for the action of $C$. For an irreducible real representation, $C$ could map the highest-weight element of the representation to the lowest-weight element, or to minus the lowest-weight element, and this fixes how $C$ acts on the entire representation.  (\ref{eq:tracekc=chi}) makes a convenient canonical sign choice. For $\mathcal{N}=4$ SYM, the overall sign for $C$ is fixed since, as explained above, $C$ acts on letters as minus the transpose. Importantly, as we will see below, the resulting action of $C$ on tensor products of letters can include an extra minus sign relative to the canonical sign choice of (\ref{eq:tracekc=chi}).}
\[ \label{eq:tracekc=chi}
\mathrm{Tr}_{\mathbf{d}} K_C(z_1,z_2 \ldots z_{[N/2]}) = 
\begin{cases}
\chi^{SO(N+1)}_{p(\mathbf{d})}(z_1, z_2 \ldots z_{N/2}) & (N \text{ even}), \\
\chi^{SP(N-1)}_{p(\mathbf{d})}(z_1, z_2 \ldots z_{(N-1)/2})  & (N \text{ odd}).
\end{cases}
\]
 Explicit expressions\footnote{We use the convention that $SP(N-1)$ has the Lie algebra $C_{(N-1)/2}$, which is consistent since characters of $SP(N-1)$ appear for $N$ odd. Also, note that since the characters of $SO(N+1)$ appear for $N$ even, this corresponds to the Lie algebra $B_{N/2}$.} for $\chi_{p(\mathbf{d})}^{SO(N+1)}(z_i)$ and $\chi_{p(\mathbf{d})}^{SP(N-1)}(z_i)$  are given in (\ref{eq:socharacter}) and (\ref{eq:spcharacterformula}).  We now recognize $p$ as a one-to-one map from Dynkin labels of real $SU(N)$ representations to Dynkin labels of orthogonal/symplectic group representations\footnote{For $N$ even, a real representation has Dynkin labels $\mathbf{d} = [d_1, d_2,\ldots d_{N/2-1}, d_{N/2}, d_{N/2-1},\ldots 1]$ so $p$ just keeps the first $N/2$ labels. The inverse map simply appends $[ d_{N/2-1},d_{N/2-2},\ldots 1]$ to $p(\mathbf{d})$. The only changes for $N$ odd follow from $d_{[N/2]}$ appearing twice in $\mathbf{d}$.}. Inspection of $K_C$ reveals that it is a real symmetric function of the $z_i$ (for $z_i = e^{i \theta_i}$), as necessary to match $\chi^{SO(N+1)}$ or $\chi^{SP(N-1)}$. Additionally, since the $\gen{J}^{2i-1}_{2 i}$ in (\ref{eq:definekc}) are mutually commuting, the exponential factor of $K_C$ factorizes into $[N/2]$ pieces that are a function of a single $z_i$. Similarly, $\chi^{SO(N+1)}$ or $\chi^{SP(N-1)}$ can be written as a trace of an orthogonal or symplectic group element that factorizes into  $[N/2]$ pieces that are a function of a single $z_i$. In the following two sections, we show that (\ref{eq:tracekc=chi}) passes many tests. For the remainder of this section we assume (\ref{eq:tracekc=chi}) holds.

Since the charged character is a trace, it satisfies the same additive and multiplicative properties of ordinary characters (\ref{eq:characterproperties}).  As for ordinary characters, for each (adjoint) letter with weight $x_B$ or $x_F$, we take into account Bose or Fermi  statistics using factors
\[ \label{eq:singlelettercontribution}
 \frac{1}{\mathrm{det}_{\text{adjoint}} \big(1 - x_B \, K_C(z_i)\big)} \quad \text{or} \quad  \mathrm{det}_{\text{adjoint}} \big(1 - x_F \, K_C(z_i)\big) .
 \]
Note that these determinants are evaluated in the adjoint representation of $SU(N)$, but due to (\ref{eq:tracekc=chi}) they expand in terms of orthogonal/symplectic group  characters. Therefore, because of the orthonormality of orthogonal/symplectic characters, the product of these factors integrated over the appropriate orthogonal or symplectic group give $\mathrm{Tr}(K_C)$ over $SU(N)$ singlet states\footnote{Here we are also using that $p$ maps Dynkin labels for  the singlet representation of $SU(N)$ to Dynkin labels for the singlet representation of the corresponding orthogonal/symplectic group.}.  On these singlet representations, we have
\[
\mathrm{Tr}_{\text{singlet}}\big(\, (-1)^F K_C(z_i) e^{\mu_i q_i}\big) = \mathrm{Tr}_{\text{singlet}}\big(\, (-1)^F C \, e^{\mu_i q_i}\big),
\]
as we need for computing the charged superconformal index. Putting all this together we have
\[ \label{eq:firstintegralequationforIc}
\mathcal{I}_C=\int_{G(N)} \mathrm{d}\mu(z_i) \,  \frac{\prod_F\mathrm{det}_{\text{adjoint}} \big(1 - x_F \, K_C(z_i)\big)}{\prod_B \mathrm{det}_{\text{adjoint}} \big(1 - x_B \, K_C(z_i)\big)}, \quad G(N) = \begin{cases} SO(N+1) & (N \text{ even}), \\ SP(N-1) & (N \text{ odd}). \end{cases}
\]
$\mathrm{d}\mu$ is the Haar measure for the group $G(N)$, and the products are over all bosonic and fermionic letters with weights $x_B$ or $x_F$. 

Straightforward evaluation of the determinants for $SU(N)$ for $N=2,3\ldots 7$ leads us to conclude that
\< \label{eq:evaluateddeterminant}
\mathrm{det}_{\text{adjoint}} \big(1 - x \, K_C\big) \eq u(x,z_i)(1-x^2)^{[N/2]} \prod_{i=1}^{[N/2]} (1 - x z_i) (1 - \frac{x}{z_i}) \prod_{1 \leq i < j \leq [N/2]} v(x,z_i,z_j),
\nln
v(x,z_i,z_j) \eq (1 - x^2 \frac{z_i}{z_j}) (1 - x^2 \frac{z_j}{z_i})(1 -x^2 z_i z_j)(1 - \frac{x^2}{z_i z_j}), 
\nln
u(x, z_i) \eq \begin{cases} (1+x)^{-1} & (N \text{ even}), \\ \prod_{i=1}^{(N-1)/2} (1 + x^2 z_i)(1 + x^2 z_i^{-1}) & (N \text{ odd}).\end{cases}
\>
See Appendix \ref{sec:appsu2su3} for the $N=2,3$ cases. The explicit calculations for  $N\leq7$ and the large-$N$ tests of Section \ref{sec:largeN} provide convincing confirmation of (\ref{eq:evaluateddeterminant}), though an analytic calculation of this determinant should be possible. This expression shows that this determinant for the charged index is very different than the determinant one would use for evaluating the ordinary index for $\mathcal{N}=4$ SYM with $G(N)$ gauge group.

We now substitute (\ref{eq:evaluateddeterminant})  into (\ref{eq:firstintegralequationforIc}) and apply this for the letters of $\mathcal{N}=4$ SYM. To simplify the resulting expression,  we write the factors of (\ref{eq:evaluateddeterminant}) in terms of exponentials of series expansions of logarithms, that is in plethystic form. Then, like the superconformal index, the matrix integral for $\mathcal{I}_C$ can be written in terms of the single-letter index $f$ (\ref{eq:n=4f}), 
\< \label{eq:matrixintegralforic}
\mathcal{I}_C (t, y, v, w) \eq \int_{G(N)} \!\! \mathrm{d}\mu(z_i) \,  \mathrm{exp} \Bigg( \sum_{m=1}^\infty \Big(  \frac{f(x^m)}{m} V_m(z_k) + \frac{f(x^{2m})}{m} W_m(z_k)  \Big) \Bigg), 
\\
f(x^m) \eq f(t^m,y^m,v^m, w^m) , \quad  G(N) = \begin{cases} SO(N+1) & (N \text{ even}), \\ SP(N-1) & (N \text{ odd}), \end{cases}
\nln
V_m(z_k) \eq   \sum_{i=1}^{[N/2]} (z_i^m+z_i^{-m}) - (-1)^m \delta^+_N , \quad \delta^\pm_N = \half (1 \pm (-1)^N),
\nln
W_m(z_k)\eq  \sum_{1 \leq i < j \leq [N/2]} (z_i^m+z_i^{-m})(z_j^m+z_j^{-m}) +  (-1)^m \delta^-_N \, V_m(z_k) + [N/2].  \notag
\>
While this expression is given as an integral over $G(N)$, the integrand shows clearly that  $\mathcal{I}_C$ is different than the index for $\mathcal{N}=4$ SYM with $G(N)$ gauge group. Also, due to the integration being over $G(N)$, there does not seem to be any advantage to considering the index of positive/negative $C$-charge states, $\half(\mathcal{I} \pm \mathcal{I}_C)$. These  positive/negative charge indices are simply linear combinations of the matrix integrals over $SU(N)$ and over $G(N)$.

It is often useful to write the matrix integrals for the superconformal index in terms of the elliptic Gamma function, defined as
\[
\Gamma(z;p,q) = \prod_{j,k\geq0} \frac{1 -z^{-1}p^{j+1}q^{k+1}}{1 - z p^j q^k}.
\]
For $\mathcal{N}=4$ SYM examples see \cite{Gadde:2009kb,Spiridonov:2010qv}. While we will not take advantage of special properties of such elliptic integrals in this work, for completeness we now write the charged index in terms of elliptic Gamma functions.
First we define the q-Pochhammer symbol,
\[
(a;b) =(a;b)_\infty = \prod_{k=0}^\infty (1 - a b^k).
\]
Also, we use a standard abbreviation for products of elliptic Gamma functions. Each factor of $z_i^\pm$ in the first argument means that we include a factor of the elliptic Gamma function using $z_i$ and another factor using $z_i^{-1}$, so for example
\<
\Gamma(\alpha z^\pm;p,q)\eq\Gamma(\alpha z;p,q)\Gamma(\alpha z^{-1};p,q), 
\nln
 \Gamma(\alpha z_i^\pm z_j^\pm;p,q) \eq \Gamma(\alpha z_i z_j;p,q)\Gamma(\alpha z_i^{-1} z_j;p,q)\Gamma(\alpha z_i z_j^{-1};p,q) \Gamma(\alpha z_i^{-1} z_j^{-1};p,q).
\>
Finally, we change variables from $(t,y,v,w)$ to $(p,q, \alpha_A)$ for $A=1,2,3$ as
\begin{align} 
p  &= t^3 y, & q  &= t^3 y^{-1},  & & \notag \\
\alpha_1 &= t^2 v, & \alpha_2 &= \frac{t^2 }{w}, & \alpha_3 &=  \frac{t^2 w }{v}.
\end{align}
Then we have for $N$ even, 
\begin{multline}
\mathcal{I}_C (\alpha,p,q) = \frac{1}{2^{N/2} (N/2)!} \frac{(p^2;p^2)^{N/2}(q^2;q^2)^{N/2}}{(-p;p)(-q;q)}\prod_{A=1}^3 \frac{\Gamma^{N/2}(\alpha_A^2;p^2,q^2)}{\Gamma(-\alpha_A;p,q)}
\\
\times \oint \prod_{i=1}^{N/2} \frac{\mathrm{d}z_i}{2 \pi i z_i} \prod_{i=1}^{N/2} \frac{\prod_{A=1}^3 \Gamma(\alpha_A z_i^\pm;p,q)}{\Gamma(z_i^\pm;p,q)}  \prod_{1 \leq i < j \leq N/2}  \frac{\prod_{A=1}^3 \Gamma(\alpha_A^2 z_i^\pm z_j^\pm;p^2,q^2)}{\Gamma(z_i^\pm z_j^\pm;p^2,q^2)}, 
\end{multline}
and for $N$ odd, 
\begin{multline} 
\mathcal{I}_C (\alpha,p,q) = \frac{1}{2^{[N/2]} [N/2]!} (p^2;p^2)^{[N/2]}(q^2;q^2)^{[N/2]}\prod_{A=1}^3 \Gamma^{[N/2]}(\alpha_A^2;p^2,q^2)
 \\
\times \oint \prod_{i=1}^{[N/2]} \frac{\mathrm{d}z_i}{2 \pi i z_i} \prod_{i=1}^{[N/2]} \frac{\prod_{A=1}^3 \Gamma(\alpha_A z_i^\pm;p,q) \Gamma(-\alpha_A^2 z_i^\pm;p^2,q^2)}{\Gamma(z_i^\pm;p,q)\Gamma(-z_i^\pm;p^2,q^2)}    \prod_{i<j}   \frac{\prod_{A=1}^3 \Gamma(\alpha_A^2 z_i^\pm z_j^\pm;p^2,q^2)}{\Gamma(z_i^\pm z_j^\pm;p^2,q^2)}. 
\end{multline}
The $z_i$ are integrated over unit circles. Also, the last product is for $1\leq i < j \leq [N/2]$.

\section{Checks for small $N$ \label{sec:smallN}}

The key equations of the previous section are the conjectures (\ref{eq:tracekc=chi}) and (\ref{eq:evaluateddeterminant}), from which the matrix integral for $\mathcal{I}_C$  (\ref{eq:matrixintegralforic}) follows.  One can directly test  these key equations for a specific $SU(N)$ irreducible representation by finding the representation matrices for the $\gen{J}^A_B$ and $C$. For example, Appendix \ref{sec:appsu2su3} verifies these equations for the adjoint representations of $SU(2)$ and $SU(3)$. In this section, we instead report on more efficient consistency tests for small $N$ that do not require us to construct the representation matrices. 

We first identify and perform a simple test of the charged character's relation (\ref{eq:tracekc=chi})  to characters of orthogonal or symplectic groups. Consider the tensor product of two real $SU(N)$ irreducible representations $R_{\mathbf{d}_1}$ and $R_{\mathbf{d}_2}$, 
\[ \label{eq:suntensorproduct}
R_{\mathbf{d}_1} \otimes R_{\mathbf{d}_2} = \sum_{\text{real }  R_{\mathbf{d}_i}} n_{12}^i R_{\mathbf{d}_i} \, \, \oplus \text{ complex representations},
\]
where  $n_{12}^i$ are the nonnegative integers  that count the number of $R_{\mathbf{d_i}}$ in the tensor product $R_{\mathbf{d}_1} \otimes R_{\mathbf{d}_2}$. Let us evaluate the charged character for the representations on both sides. For the left side, assuming (\ref{eq:tracekc=chi}), the charged character is the product of characters of $G(N) = SO(N+1)$ or $SP(N-1)$ for $N$ even or odd respectively. For the right side, we have separated out complex representations (which come in complex conjugate pairs) since they do not contribute to the charged character. Importantly, for the real $R_{\mathbf{d}_i}$ appearing in the sum, the charged character can be plus or minus the character of the  $G(N)$ representation $R_{p(\mathbf{d}_i)}$. This is because the action of $C$ on the $R_{p(\mathbf{d}_i)}$ follows from its action on $R_{\mathbf{d}_1}$ and $R_{\mathbf{d}_2}$, and need not match the canonical sign choice of (\ref{eq:tracekc=chi}). So we have
\[
\chi^{G(N)}_{p(\mathbf{d_1})} \cdot \chi^{G(N)}_{p(\mathbf{d_2})} = \sum_{i} m^{i}_{12} \chi^{G(N)}_{p(\mathbf{d_i})}, \quad   n^{i}_{12} \geq |m^{i}_{12}|  \, \text{ and } \,  n^{i}_{12}= m^{i}_{12} \, \, \,  \mathrm{mod} \,  2,
\]
where the summation is over the same $i$ as in (\ref{eq:suntensorproduct}). On the other hand, the left side gives the character for the tensor product of $G(N)$ representations\footnote{Recall that (\ref{eq:definep}) implies that $p$ is a one-to-one map between  $SU(N)$ real irreducible representations and $G(N)$ representations, so we write the Dynkin labels for a $G(N)$ representation uniquely as $p$ of the Dynkin labels of a real $SU(N)$ representation.},
\[
\chi^{G(N)}_{p(\mathbf{d_1})} \cdot \chi^{G(N)}_{p(\mathbf{d_2})} = \sum_{i} l^{i}_{12} \chi^{G(N)}_{p(\mathbf{d_i})},
\]
where  $l^{i}_{12}$ are the nonnegative integers that give the number of $R_{p(\mathbf{d_i})}$ in the $G(N)$ tensor product $R_{p(\mathbf{d_1})}\otimes R_{p(\mathbf{d_2})}$. As before, the summation runs over all $i$ such that $R_{\mathbf{d_i}}$ is a $SU(N)$ real irreducible representation.
The last two equations imply that $l^{i}_{12} = m^{i}_{12}$, and this gives relations between $SU(N)$ and $G(N)$ tensor product coefficients $n_{12}^i$ and $l_{12}^i$, 
 \[ \label{eq:su(n)g(n)relation}
  n^{i}_{12} \geq l^{i}_{12}  \, \text{ and } \,  n^{i}_{12}= l^{i}_{12} \, \, \,  \mathrm{mod} \,  2.
  \]
Using the computer algebra program LiE \cite{Leeuwen:1992xx}, we have verified (\ref{eq:su(n)g(n)relation}) for\footnote{The relation is trivial for $N=2$ since the Lie algebras of $SU(2)$ and $SO(3)$ are isomorphic, which matches the fact that all $SU(2)$ representations are real.}  $3 \leq N \leq 7$   for random representations with Dynkin label components less than ten.

For a specific example, consider the tensor product of $SU(5)$ representations with Dynkin labels $[1,2,2,1]$ and $[0,1,1,0]$. This tensor product expands as 
\begin{multline}
[1,2,2,1] \otimes [0,1,1,0] = [1,3,3,1] \oplus 2[2,2,2,2]  \oplus [3,1,1,3] \oplus 2[0,3,3,0] \\
\oplus 5[1,2,2,1] \oplus2[2,1,1,2]  \oplus 2[0,2,2,0]  \oplus[1,1,1,1] \oplus \text{complex representations}.
\end{multline}
Dropping the complex representations and acting with the projector $p$ on both sides, (\ref{eq:su(n)g(n)relation}) implies for a $SP(4)$ or $C_2$ tensor product,
\[ \label{eq:mod2}
[1,2] \otimes [0,1] = [1,3] \oplus 2[2,2]  \oplus [3,1] \oplus 2[0,3]  \oplus 5[1,2] \oplus 2[2,1]  \oplus 2[0,2]  \oplus[1,1] \mod 2, 
\]
where ``$\mathrm{mod} \, 2$'' refers to the coefficients of the representations (Dynkin labels). In fact, the actual $SP(4)$ tensor product is
\[
[1,2] \otimes [0,1] = [1,3] \oplus [3,1] \oplus [1,2] \oplus [1,1] , 
\]
which is in agreement ($\mathrm{mod} \, 2$)  with (\ref{eq:mod2}).

We now  check the relation to orthogonal/symplectic characters (\ref{eq:tracekc=chi}) further while also checking our expressions for the determinants (\ref{eq:evaluateddeterminant}). Let us compare the contribution from a single bosonic or fermionic letter to the superconformal index (\ref{eq:detineigenbasis},  \ref{eq:fermionicexpform}) to the single-letter contribution to the charged index (\ref{eq:singlelettercontribution}). First, the single-letter contribution to the superconformal index expands in $SU(N)$ characters as 
\[ \label{eq:singlelettersuncontribution}
\mathrm{det}_{\text{adjoint}}^\pm (1 - x K(z_i))= \sum_{\text{real } \mathbf{d}} \sum_{j=0}^\infty n^\pm_{j,\, \mathbf{d}} x^j \chi_{\mathbf{d}}(z_i) \, \, \oplus  \, \text{complex representations}.
\]
For fermionic letters (positive power of the determinant), the $n^+_{j,\, \mathbf{d}}$ can be negative. It is convenient to package this expansion in a generating function. For that purpose, we define the generating function for characters of real irreducible representations of $SU(N)$, 
\[
\Phi(a_i;z_i) = \sum_{\text{real }\mathbf{d}} \chi_{\mathbf{d}}( z_i)  \prod_{k=1}^{[N/2]} a_k^{d_k}.
\]
For $N$ even we restrict the sum to representations that have an even last Dynkin label $d_{N/2}$, since only these representations appear in (\ref{eq:singlelettersuncontribution}). Then the generating function for the  $n^\pm_{j,\, \mathbf{d}}$ is
\[
I^{\pm}(x; a_i) = \int_{SU(N)} \mathrm{d}\mu(z_i)\, \mathrm{det}_{\text{adjoint}}^\pm (1 - x K(z_i))\,\Phi(a_i;z_i) = \sum_{\text{real } \mathbf{d}} \sum_{j=0}^\infty n^\pm_{j,\, \mathbf{d}} x^j \prod_{k=1}^{[N/2]} a_k^{d_k}.
\]
The second equality follows from (\ref{eq:singlelettersuncontribution}) and the orthogonality of $SU(N)$ characters\footnote{Note that $\Phi(a_i;z_i)=\Phi^\ast(a_i;z_i)$.}. We see that $n^\pm_{j,\, \mathbf{d}}$ is the coefficient of $x^j \prod a_k^{d_k}$ of $I^{\pm}(x; a_i)$.

Similarly, assuming (\ref{eq:tracekc=chi}), we have
\[ \label{eq:singlelettergncontribution}
\mathrm{det}_{\text{adjoint}}^\pm (1 - x K_C(z_i))= \sum_{\text{real } \mathbf{d}} \sum_{j=0}^\infty m^\pm_{j,\, p(\mathbf{d})} x^j \chi^{G(N)}_{p(\mathbf{d})}(z_i) .
\]
Now using orthogonality of $G(N)$ characters, we can define a generating function for the $m^\pm_{j,\, p(\mathbf{d})}$ as
\begin{gather}
\Theta^{G(N)}(a_i;z_i) = \sum_{\text{real }\mathbf{d}} \chi^{G(N)}_{p(\mathbf{d})}( z_i)  \prod_{k=1}^{[N/2]} a_k^{d_k},
\\
I^{\pm}_C(x; a_i) = \int_{G(N)} \mathrm{d}\mu(z_i) \, \mathrm{det}_{\text{adjoint}}^\pm (1 - x K_C(z_i))\,\Theta^{G(N)}(a_i;z_i) = \sum_{\text{real } \mathbf{d}} \sum_{j=0}^\infty m^\pm_{j,\, \mathbf{d}} x^j \prod_{k=1}^{[N/2]} a_k^{d_k}. \notag
\end{gather}
As for $\Phi$, for $\Theta^{G(N)}$ for even $N$ we only sum over even $d_{N/2}$ since only representations of this type (non-spinor representations of $SO(N+1)$) appear in (\ref{eq:singlelettergncontribution}). By the same reasoning as we used leading to (\ref{eq:su(n)g(n)relation}),  (\ref{eq:tracekc=chi}) implies that
\[ \label{eq:nmcondition}
|n^\pm_{j,\, \mathbf{d}}| > |m^\pm_{j,\,   p(\mathbf{d})}|  \, \text{ and } \,  |n^\pm_{j, \, \mathbf{d}}| = |m^\pm_{j,  \, p(\mathbf{d})}| \, \, \,  \mathrm{mod} \,  2,
\]
where we need absolute values for the $n^\pm$ only for the possibly negative fermionic coefficients $n^+_{j, \, \mathbf{d}}$.
By computing the generating functions for  $n^\pm$ and  $m^\pm$, we have checked this for $N=2,3,4,5$, and in all cases (\ref{eq:nmcondition}) is satisfied. Below we summarize the computation for the $N=2,3$ cases, and Appendix \ref{sec:appdata} gives the results for $N=4,5$.

First we discuss some ways that we simplify the calculation of $I^\pm(x;a_i)$. In our expression for $\Phi$ for $SU(N)$, we  substitute $N!$ times one of the $N!$ terms in the determinant in the numerator of the character formula (\ref{eq:suncharacter}), since all other terms just follow from permutations of the $z_i$, and the remaining factors in the integrand are symmetric under these permutations. Also, $\Phi$  can be written more compactly using the Vandermonde determinant $\Delta(z)$,
\[
\Delta(z) = \prod_{1 \leq i < j \leq N} (z_i - z_j).
\]
Instead of integrating over  $SU(N)$ using a $\delta (1-\prod z_i )$ factor and the $U(N)$ Haar measure, we write the integrand as a homogeneous function of the $N$ $z_i$, that is a function which is invariant under multiplying all $z_i$ by a common factor. As shown in Appendix \ref{sec:appformulas}, by integrating with the symmetric measure  (\ref{eq:sunmeasure})  we obtain the correct result for the integral. With this method, the $SU(N)$ integrals can be done easily by residues, using $|a_i| < 1$ and $|x| < 1$. Finally,  we use the explicit formula
\[ \label{eq:formuladetsun}
\mathrm{det}_{\text{adjoint}}(1 - x K(z_i))= (1-x)^{N-1}\prod_{1 \leq i < j \leq N} (1 - x \frac{z_i}{z_j})(1 - x \frac{z_j}{z_i}).
\]

For the calculation of $I_C^\pm(x;a_i)$, symmetry allows us to replace $[N/2]!$ terms in $\Theta^{G(n)}$ with one term from the numerator of the character formulas for $G(N)$ (\ref{eq:socharacter}) and (\ref{eq:spcharacterformula}). Additionally, symmetry under $z_i \rightarrow z_i^{-1}$ allows us to use twice the positive exponent of the $z_{[N/2]}$ factor from the numerator of the character formulas, dropping the negative exponent contribution. Now the measures are given in (\ref{eq:someasure} - \ref{eq:spmeasure}), and we use the determinant formula  (\ref{eq:evaluateddeterminant}), which is being tested. Again, we can do the integrations by residues  using $|a_i| < 1$ and $|x| < 1$.

Applying the above for $N=2$, first we compute the generating function for characters, 
\[
\Phi(a_1;z_1,z_2) = \sum_{d_1 = 0, \text{ even} }^\infty  a_1^{d_1} \chi_{[d_1]}( z_1,z_2) \rightarrow 2! \frac{z_1 z_2}{(z_1-z_2)(z_2 - a_1^2 z_1)}.
\]
Then we multiply by the inverse of the determinant  (\ref{eq:formuladetsun})  and integrate over $SU(2)$ (for which $\Delta= (z_1-z_2)$), obtaining
\<
I^{-}(x,a_1) \eq \frac{1}{2} \oint \!\frac{\mathrm{d}z_1}{2 \pi i z_1} \!\oint \!\frac{\mathrm{d}z_2}{2 \pi i z_2} (z_1-z_2)(\frac{1}{z_1}-\frac{1}{z_2})  \frac{2 z_1 z_2}{(z_1-z_2)(z_2 - a_1^2 z_1)}  \frac{(1 - x)^{-1}}{(1 - x \frac{z_1}{z_2})(1 - x \frac{z_2}{z_1})}
\nln
\eq   \oint \!\frac{\mathrm{d}z_1}{2 \pi i } \!\oint \!\frac{\mathrm{d}z_2}{2 \pi i }   \frac{ (z_2-z_1)}{(z_2 - a_1^2 z_1)}  \frac{(1 - x)^{-1}}{(z_2 - x \, z_1)(z_1 - x \, z_2)}
\nln
\eq    \!\oint \!\frac{\mathrm{d}z_2}{2 \pi i }   \frac{ (z_2-x \, z_2)}{(z_2 - a_1^2 x z_2)}  \frac{(1 - x)^{-1}}{z_2 - x^2 z_2}
\nln
\eq  \frac{1}{(1-x^2)(1 - a_1^2 x)}.
\>
The last two lines follow from evaluating the residues at $z_1=x \, z_2$ and then at $z_2=0$. A similar calculation for the fermionic case of $I^{+}$ gives
\[
I^{+}(x;a_1)=1 - x \, a_1^2+ x^2 a_1^2 - x^3.
\]
Next for $G(2) = SO(3)$,  we compute the generating function for characters, 
\[
\Theta^{SO(3)}(a_1;z) =\sum_{d_1 \text{ even}}  a_1^{d_1} \chi_{[d_1]}^{SO(3)}(z) \rightarrow -2  \frac{z}{(1-z)(1-a_1^2z)}.
\]
Then we multiply by the inverse of the determinant (\ref{eq:evaluateddeterminant}) and integrate over $SO(3)$ using the measure (\ref{eq:someasure}),  obtaining
\<
I_C^{-}(x,a_1) \eq -\frac{1}{2} \oint \frac{(z+z^{-1}-2)\mathrm{d}z}{2 \pi i z}  (-2)  \frac{z}{(1-z)(1-a_1^2z)}  
\frac{1+x}{(1-x^2)(1 - x \, z)(1 - \frac{x}{z})}
\nln
\eq \frac{1}{(1-x^2)(1 - a_1^2 x)}.
\>
Evaluating the residue at $z=x$ and simplifying yields the second line. Again repeating for the fermionic generating function we find
\[
I_C^{+}(x;a_1)=1 - x \, a_1^2+ x^2 a_1^2 - x^3.
\]
Since $I^{\pm}_C(x;a_1) = I^{\pm}(x;a_1)$, the $N=2$ case is (trivially) consistent with (\ref{eq:nmcondition}), with the precise equality due to the fact that all $SU(2)$ representations are real.

Repeating the same steps for $SU(3)$ we compute 
\< \label{eq:SU3check}
\Phi(a_1; z_1, z_2, z_3) \eq \sum_{d_1 = 0 }^\infty  a_1^{d_1} \chi_{[d_1,  d_1]}( z_1,z_2, z_3) \rightarrow 3! \frac{z_1^2 z_2 z_3}{\Delta(z)(z_3-a_1 z_1 )},
\nln
 I^-(x, a_1) \eq \frac{1}{(1 - x^2)(1 - x^3)(1 - a_1 x)(1 - a_1 x^2)},
\nln
 I^+(x, a_1) \eq (1 + x^8) - (x+x^7) a_1 + (x^2 + x^6) a_1 
 \nl
 - (x^3 + x^5)(1 + a_1 + a_1^2) + x^4(2 a_1 + 2 a_1^2). 
\>
We need to compare the last two lines to $I_C^\pm$ for $G(3) = SP(2)$. We find %
\<  \label{eq:G3check}
\Phi^{SP(2)}(a_1, z_1) \eq  \sum_{d_1}  a_1^{d_1}\chi^{C_1}([d_1], z_1)  \rightarrow -2  \frac{z_1^2}{(1-a_1 z_1)(1-z_1^2)}, 
\nln
I_C^-(x, a_1) \eq  \frac{1}{(1-x^2)(1+x^3)(1 - a_1 x)(1+ a_1 x^2)},
\\
I_C^+(x, a_1)  \eq  (1-x^8) - (x-x^7) a_1 + (x^2 -x^6)a_1 - (x^3-x^5) (1 - a_1 + a_1^2). \notag 
\>
Reading off the expansion coefficients $n^\pm_{j, \, \mathbf{d}}$ and $m^\pm_{j, \, p(\mathbf{d})}$ from (\ref{eq:SU3check}) and (\ref{eq:G3check}), we see that  (\ref{eq:nmcondition}) is satisfied. 
Appendix \ref{sec:appdata} reports the results of the similar calculations for $N=4,5$, where the the $I^\pm$ and $I^\pm_C$ are given by lengthier expressions but still are consistent with (\ref{eq:nmcondition}).
\section{Checks at large $N$ \label{sec:largeN}}

Below we will use the saddle-point method to evaluate $\mathcal{I}_C$ in the large-$N$ limit. Then we will show that this matches the planar limit of $\mathcal{I}_C$.

A useful simple example is given by the charged index (or partition function) for a single scalar $SU(N)$ adjoint field $\phi$, with weight $x$. In the planar limit, the single-trace charged index becomes %
\[ \label{eq:singlescalarsingletraceindex}
\mathcal{I}_{C,\text{ s.t.}}(x) = \frac{x^2}{1 + x}.
\]
This corresponds to the states $\mathrm{Tr}(\phi^n)$ for $n \geq 2$, including the fact that $C$ acts as $(-1)^n$ in the planar limit. The full (multitrace) planar charged index is then
\[ \label{eq:singlescalarplanarindex}
\mathcal{I}_C(x) = \prod_{i=2}^\infty \frac{1}{1 -(-1)^i x^i}= \frac{1}{(x^2;-x)}.
\]
We will confirm that the results below reduce to this simple expression when the only letter is $\phi$, that is when then single-letter index is $f=x$ \footnote{Another simple check that could be done is for a single fermion with $f=-x$ (and $f(x^m) = -x^m$). In this case, $\mathcal{I}_{C,\text{ s.t.}}(x) = \frac{x^3}{1 + x^2}$, and $\mathcal{I}_C(x)=1/(x^3;-x^2)$.}.

\subsection{The large-$N$ limit of the matrix integral}

To  evaluate the large-$N$ limit it is convenient to use  the integration variables $\theta_i$ defined by $e^{i \theta_j} = z_j$. In these variables, the matrix integral for $\mathcal{I}_C$ (\ref{eq:matrixintegralforic}) becomes,
\< \label{eq:icexpressioninangularcoordinates}
\mathcal{I}_C (t, y, v, w) \eq \int_{G(N)} \!\! \mathrm{d}\mu(\theta_i) \,  \mathrm{exp} \Bigg( \sum_{m=1}^\infty \Big(  \frac{f(x^m)}{m} V_m(\theta_k) + \frac{f(x^{2m})}{m} W_m(\theta_k)  \Big) \Bigg), 
\\
V_m(\theta_k) \eq   \sum_{i=1}^{[N/2]} 2 \cos (m  \, \theta_i) - (-1)^m \delta^+_N , \quad \delta^\pm_N = \half (1 \pm (-1)^N),
\nln
W_m(\theta_k)\eq  \sum_{1 \leq i < j \leq [N/2]} 4 \cos ( m \, \theta_i) \cos ( m \, \theta_j) +  (-1)^m \delta^-_N \, V_m(\theta_k) + [N/2].  \notag
\>
$\mathrm{d}\mu(\theta_i)$, the Haar measure for the orthogonal or symplectic group $G(N)$,  is given in (\ref{eq:someasure} - \ref{eq:spmeasure}).

We will evaluate the limit for even $N$. The odd-$N$ calculation proceeds similarly. For $t,y,v,w=0$ (which we abbreviate as $x=0$), $\mathcal{I}_C$ is 1 for all $N$. Therefore, we will drop some terms that are independent of the chemical potentials $x$ and of $\theta_i$, restoring the correct normalization at the end of the calculation.   Absorbing factors from the measure into an exponential,  for even $N$ $\mathcal{I}_C$ becomes
 \<  \label{eq:expressionsforVi}
 \mathcal{I}_C(t,y,v,w) & \propto & \int \mathrm{d}\theta_i \exp \big(-\sum_{i \neq j} V_2(\theta_i, \theta_j) -\sum_{i } V_1(\theta_i) - V_0 \big),
 \nln
 V_2(\theta_1, \theta_2) \eq \sum_{m=1}^\infty  \frac{1 - f(x^{2m})}{m} 2 \cos (m \, \theta_1) \cos (m \, \theta_2),
 \nln
 V_1((\theta) \eq   \sum_{m=1}^\infty \frac{ 1-f(x^{m})}{m}  2  \cos (m \, \theta),
 \nln
 V_0 \eq  -\sum_{m=1}^\infty \frac{N}{2} \frac{ f(x^{2m})}{m} + (-1)^m \frac{f(x^m)}{m}.
 \>
To obtain this we used the following series expansions,
\<
\log ((\cos \theta_1-\cos \theta_2)^2) \eq - \log 4 - \sum_{n=1}^\infty \frac{4}{n} \cos (n \, \theta_1) \cos ( n \, \theta_2),
\nln
\log \sin^2(\theta/2) \eq  - \log 4 - \sum_{n=1}^\infty \frac{2}{n} \cos ( n \, \theta). 
\>
As usual, to evaluate the large-$N$ limit we replace the integral over the $\theta_i$ with an integral over the Fourier modes of the eigenvalue density $\rho$. The eigenvalue density is normalized so that
 \[
 \int_0^{2 \pi} \mathrm{d} \theta \rho(\theta) = 1.
 \]
 We normalize the Fourier modes as
 \[
 \rho_n = \int_0^{2 \pi} \mathrm{d} \theta \rho(\theta) \cos (n \, \theta).
 \]
 To find the effective potential in terms of the Fourier coefficients we need to evaluate
 \[
 \Big(\frac{N}{2} \Big)^2 \int \mathrm{d} \theta_1 \mathrm{d} \theta_2 \, \rho(\theta_1) \rho(\theta_2) V_2 (\theta_1, \theta_2) + \frac{N}{2}    \int \mathrm{d} \theta \rho(\theta) \Big(-V_2^{\text{reg.}} (\theta) +  V_1 (\theta) \Big)+ V_0.
 \]
At $\mathcal{O}(N)$, in addition to  $V_1$,  there is an extra contribution  to cancel the $\mathcal{O}(N)$ contributions from $V_2(\theta_i,\theta_j)$ for $i=j$, which are not included in $\mathcal{I}_C$ (\ref{eq:icexpressioninangularcoordinates}). However, since $V_2$ diverges when its arguments are equal, we define the regularized potential for equal arguments as
\[
 V_2^{\text{reg.}}(\theta) = \sum_{m=1}^M  \frac{1 - f(x^{2m})}{m} 2 \cos^2 ( m \, \theta ),
\]
for some large $M$. Note that for $M=\infty$,   $V_2^{\text{reg.}}(\theta) =  V_2(\theta, \theta)$ (formally). As we will see, the divergence for $M \rightarrow \infty$ can be canceled by removing a term that is independent of $x$ and $\rho$, so this regularization allows us to find the large-$N$ limit up to overall normalization.   Substituting the expressions for the $V_i$ (\ref{eq:expressionsforVi}) and simplifying we obtain
 \<
V(\rho,M) \eq  \Big(\frac{N}{2} \Big)^2 \sum_{m=1}^\infty  \frac{1 - f(x^{2m})}{m} 2 \rho_m^2 - \frac{N}{2}   \sum_{m=1}^M  \frac{1 - f(x^{2m})}{m} (\rho_{2m}+1) 
\nl
+\frac{N}{2}   \sum_{m=1}^\infty \frac{1- f(x^{m})}{m} 2 \rho_m  -  \sum_{m=1}^\infty \frac{N}{2}   \frac{ f(x^{2m})}{m} + (-1)^m \frac{f(x^m)}{m}
\nln
\eq
 \Big(\frac{N}{2} \Big)^2 \sum_{m=1}^\infty  \frac{1 - f(x^{2m})}{m} 2 \rho_m^2  + \frac{N}{2} \sum_{m=1,\text{ odd}}^{2M} \frac{1 - f(x^{m})}{m} 2 \rho_m   
  \\
  & &  +\frac{N}{2}   \sum_{m=2M+1}^\infty \frac{1- f(x^{m})}{m} 2 \rho_m  -  \frac{N}{2}  \sum_{m=M+1}^\infty   \frac{ f(x^{2m})}{m}   - \frac{N}{2} \sum_{m=1}^M  \frac{1}{m} 
  \nl
  +  \sum_{m=1}^\infty  (-1)^m\frac{ f(x^m)}{m}. \notag
  \>
Dropping the $\rho$- and $x$-independent second-to-last term, and then sending $M$ to infinity, yields the regularized effective potential of
\[
V(\rho) =  \Big(\frac{N}{2} \Big)^2 \sum_{m=1}^\infty  \frac{1 - f(x^{2m})}{m} 2 \rho_m^2  + \frac{N}{2} \sum_{m=1,\text{ odd}}^\infty \frac{1 - f(x^{m})}{m} 2 \rho_m   +  \sum_{m=1}^\infty  (-1)^m\frac{ f(x^m)}{m}.
\]
 Since $1-f$ (\ref{eq:1minusf}) is positive for all allowed chemical potentials, like the superconformal index, $\mathcal{I}_C$  has only a single saddle point. Here,   $\rho_m$ is $\mathcal{O}(N^{-1})$ at the saddle point. Integrating over the $\rho_m$ using the elementary Gaussian integral
\[
\int_{-\infty}^\infty \mathrm{d} \rho \, e^{\alpha \rho^2 + \beta \rho} \propto \frac{e^{-\beta^2/(4 \alpha)}}{\sqrt{- \alpha}}
\]
and normalizing as stated above so that $\mathcal{I}_C(0)=1$, we finally obtain
\< \label{eq:largenlimitic}
\lim_{N \rightarrow \infty} \mathcal{I}_C(t,y,v,w) \eq \prod_{m=1}^\infty\frac{1}{(1 - f(x^{2m}))^{1/2}}
\nl
\times  \exp \Big(\sum_{m=1, \text{ odd}}^\infty \frac{(1 - f(x^m))^2}{2 m (1 - f(x^{2m}))} - \frac{1}{2 m} \Big) \exp \Big (\sum_{m=1 }^\infty (-1)^{m+1} \frac{f(x^m)}{m} \Big).
\nl
\>
The similar calculation for $N$ odd also yields (\ref{eq:largenlimitic}).

As stated at the beginning of this section, we now confirm that for $f=x$ (instead of the $\mathcal{N}=4$ SYM single-letter index) (\ref{eq:largenlimitic}) reduces to the large-$N$ charged index for a single adjoint scalar (\ref{eq:singlescalarplanarindex}). Setting $f=x$ and simplifying the second factor gives
\< \label{eq:partlysimplified}
\lim_{N \rightarrow \infty} \mathcal{I}_C(x) \eq \prod_{m=1}^\infty\frac{1}{(1 - x^{2m})^{1/2}}  \exp \Big(\sum_{m=1, \text{ odd}}^\infty -\frac{x^m}{ m (1 + x^{ m})}  \Big) \exp \Big (\sum_{m=1 }^\infty (-1)^{m+1} \frac{x^m}{m} \Big).
\nl
\>
For the second factor, we expand the summand  as a geometric series, and also replace the sum over odd $m$ by a sum over all $m$ with a factor of $\half(1 -(-1)^m)$. We then find that
\[
 \exp \Big(\sum_{m=1, \text{ odd}}^\infty -\frac{x^m}{ m (1 + x^{ m})}  \Big)  = \sqrt{\frac{(x;x^2)(-x^2;x^2)}{(x^2;x^2)(-x;x^2)}} = \sqrt{\frac{(x;-x)}{(-x;-x)}}.
 \]
 The first factor of (\ref{eq:partlysimplified})  is $1/\sqrt{(x^2;x^2)}=1/\sqrt{(x;-x)(-x;-x)}$ and the last factor is $(1+x)= (-x;-x)/(x^2,-x)$. Combining the three factors finally yields
 \[
 \lim_{N \rightarrow \infty} \mathcal{I}_C(x) = \frac{1}{(x^2;-x)},
 \]
matching (\ref{eq:singlescalarplanarindex}).
\subsection{The charged index in the planar limit from Polya counting}

In this section, we will first compute $\mathcal{I}_C$ on single-trace states in the planar limit.  After comparing the result to the supersymmetric partition function for single-trace states \cite{Janik:2007pm}, we will evaluate the complete planar charged index that includes general multi-trace states.  We will see that this planar charged index  equals the large-$N$ limit (\ref{eq:largenlimitic}) of the proposed matrix integral for $\mathcal{I}_C$.

Counting the single-trace states is equivalent to counting words that are built from the $\mathcal{N}=4$ SYM letters and that are identified under cyclic permutations. For $\mathcal{I}_C$ this counting needs to be graded by  fermion number and $C$ charge. As explained in Section \ref{sec:computingchargedindex},  charge conjugation acts on each letter by giving minus the transpose. Therefore, in the planar limit, charge conjugation reverses the order of letters inside each word and multiplies by $(-1)^L$, where $L$ is the length of the word.  Words that are mapped to plus/minus themselves by $C$ contribute to the single-trace planar $\mathcal{I}_C$, while those that are mapped to a distinct word do not contribute (since we can make a $\pm$ doublet of states from the original word and its image under $C$). 

To count these states, we use Polya's method, which was previously applied to cyclic states in gauge theory in \cite{Sundborg:1999ue,Polyakov:2001af}.  Polya's method applies to words built from a set of weighted letters  that are invariant under a subgroup $G$ of the permutation group $S_k$. We introduce the sum of the weights
\[
z(x) = \sum_{\alpha} x_{\alpha},
\]
where $\alpha$ runs over the letters and the $x_{\alpha}$ are the weights. In our case  $x_{\alpha}$ will include the superconformal charges, fermion number, and an extra minus sign to account for the $(-1)^L$ mentioned above, so we will later substitute $-f(x)=-f(t,y,v,w)$ for $z(x)$.  According to Polya's theorem, the number of words $Z^{(G)}(k, z(x))$ of length $k$ built from the weighted letters is
\[ \label{eq:polya}
Z^{(G)}(k, z(x)) =\frac{1}{|G|}  \sum_{l=1}^{|G|} \Big( \prod_{i=1}^k z(x^i)^{n(k,g_l)_i}  \Big).
\]
Here the sum is over the elements of $G$ labeled $g_l$, and $n(k, g_l)_i$ is the number of cycles of length $i$ in the permutation $g_l$ of $k$ elements.  The permutation subgroup that includes reflections and cyclic permutations is the Dihedral group $D_k$. Words that are mapped to themselves by $C$ are counted once both when we identify words under $D_k$ and when we  identify words only under the cyclic permutation group $C_k$. However, pairs of words that are mapped to each other by charge conjugation count as one  word for $D_k$ and as two words for $C_k$. Therefore, the single-trace contribution for length $k$ is 
\[
Z^{(\text{s.t.})}(k, z(x)) = 2 Z^{D_k}(k, z(x)) - Z^{C_k}(k, z(x)) .
\]
Applying (\ref{eq:polya}), we have
\[ 
Z^{D_k}(k, z(x)) = \frac{1}{2 k}  \sum_{l=1}^k \Big( \prod_{i=1}^k z(x^i)^{n(k,l)_i} + \prod_{i=1}^k z(x^i)^{n'(k,l)_i}  \Big).
\]
$n(k,l)_i$ gives the number of cycles of length $i$ in a cyclic shift by $l$ sites of $k$ objects, and $n'(k,l)_i$ gives the number of cycles of length $i$ in a reflection followed by a shift by $l$ sites of $k$ objects. The first term of the summand corresponds to the cyclic subgroup $C_k$ of the dihedral group. It follows that $Z^{(\text{s.t.})}(k, z(x))$ is given by twice the second term,
\[
Z^{(\text{s.t.})}(k, z(x)) =  \frac{1}{ k}  \sum_{l=1}^k \Big(  \prod_{i=1}^k z(x^i)^{n'(k,l)_i}  \Big).
\]
So we need to evaluate $n'(k,l)_i$.  

Consider the action of a dihedral group element on $(1,2, \ldots k)$. After a reflection  and shift by $l$, $j$ is now at position $(k+1-j+l \mod k)$ \footnote{We write $0 \mod k$ as $k$ rather than 0.}. Since $(k+1-j+l \mod k)$ is mapped to $j$,  generically we have the two-cycles $(j, (k+1-j+l \mod k))$. So $n'(k,l)_i=0$ for $i>2$. However, when $j= k+1-j+l \mod k$, this instead becomes a one-cycle.  So the number of one cycles $n'(k,l)_1$ is given by the number of $j$ between $1$ and $k$ such that  $j= k+1-j+l \mod k$.  It is straightforward to find that this depends only on the parity of $k$ and $l$ as 
\[
n'(k,l)_1 = 
\begin{cases}
0 &: \quad \text{$k,l$ even,} \\ 2 &: \quad \text{$k$ even, $l$ odd,}  \\ 1 &: \quad \text{$k$ odd.} 
\end{cases}
\]
It follows that 
\[
n'(k,l)_2 = 
\begin{cases}
k/2 &: \quad \text{$k,l$ even,} \\ k/2-1 &: \quad \text{$k$ even, $l$ odd,}  \\ (k-1)/2 &: \quad \text{$k$ odd.} 
\end{cases}
\]
For even $k$ we then have
\<
Z^{(\text{s.t.})}(k, z(x))_{\text{$k$ even}} \eq  \frac{1}{k}\sum_{l=1}^k \Big(  z(x)^{n'(k,l)_1} z(x^2)^{n'(k,l)_2} \Big)
\nln
\eq  \frac{1}{2 k}\sum_{l=1}^k \Big( (1 + (-1)^L) z(x^2)^{k/2} + (1 - (-1)^L) z(x)^2 z(x^2)^{k/2-1} \Big)
\nln
\eq  \frac{1}{2} \big( z(x^2)^{k/2} + z(x)^2 z(x^2)^{k/2-1}  \big). 
\>
Similarly, for odd $k$,
\<
Z^{(\text{s.t.})}(k, z(x))_{\text{$k$ odd}} \eq  \frac{1}{k}\sum_{l=1}^k \Big(  z(x)^{n'(k,l)_1} z(x^2)^{n'(k,l)_2} \Big)
\nln
\eq  \frac{1}{ k}\sum_{l=1}^k \Big(z(x) z(x^2)^{(k-1)/2} \Big)
\nln
\eq  z(x) z(x^2)^{(k-1)/2} . 
\>
Now we sum over the length $k$ from one to infinity. For $SU(N)$, the $k=1$ contribution is absent, but we temporarily include this single-letter contribution for comparison to the $U(N)$ supersymmetric partition function. We will subtract the $k=1$ contribution below. The sum simplifies as
\<
Z^{(\text{s.t.})}( z(x))\eq \sum_{k=1}^\infty Z^{(\text{s.t.})}(k, z(x))
\nln
\eq \frac{1}{2}\frac{z(x^2
) + z(x)^2}{1- z(x^2)} + \frac{z(x)}{1 - z(x^2)}
\nln
\eq \frac{(z(x)+1)^2
 }{2(1- z(x^2))} -\frac{1}{2}.
\>

As explained above, to obtain $\mathcal{I}_{C, \text{ s.t.}}$ we  substitute $-f(x)$ for $z(x)$,
\[ \label{eq:icst}
\mathcal{I}_{C, \text{ s.t.}} = \frac{(1-f(x))^2
 }{2(1- f(x^2))} -\frac{1}{2}.
\]
As a check,  for the single scalar $\phi$ case of $f=x$ 
\[
\mathcal{I}_{C, \text{ s.t.}}(x) = -\frac{x}{1+x},
\]
which agrees with the expression we found earlier (\ref{eq:singlescalarsingletraceindex})
except that here $\mathrm{Tr}(\phi)$ is included since we are temporarily considering a $U(N)$ rather than $SU(N)$ theory. Returning to $\mathcal{N}=4$ SYM, we substitute the expression for $f$ (\ref{eq:n=4f}) and find
\[
\mathcal{I}_{C, \text{ s.t.}} = \frac{(y + \frac{1}{y} ) (t^3 +  t^7 w  + \frac{t^7}{v}  + \frac{t^7 v }{w} ) - (t^2 + 
    t^8) (v  + \frac{1}{w} +  \frac{w}{v} + t^4 )  }{(1 + t^2 v)(1 + t^2/w)(1 + t^2 w/v)(1 - t^3 y)(1 - t^3/y)}.
    \]
It is interesting to compare this to the partition function of $1/16$ BPS states. Janik and Trzetrzelewski \cite{Janik:2007pm} wrote the partition in terms of oscillators  \cite{Gunaydin:1984fk,Beisert:2003jj}. To translate their expressions into the variables of the index use the following dictionary,
\begin{align} \label{eq:oscillatortochemicalpotentials}
a_2 & =  t^3, & b_1 & =  y, & b_2 & =  \frac{1}{y},
\notag \\
c_2 & =  \frac{t}{v}, & c_3 & =  \frac{ t v}{w}, &  c_4 & = t w.
\end{align}
In these variables, the single-trace partition function for a $U(N)$ theory, which is given by equation (41) of \cite{Janik:2007pm}, simplifies to
\[ \label{eq:susypartition}
Z_{\text{s.t.}}^{1/16th}(t, v, w, y) = \frac{(y + \frac{1}{y} ) (t^3 +  t^7 w  + \frac{t^7}{v}  + \frac{t^7 v }{w} ) + (t^2 + 
    t^8) (v  + \frac{1}{w} +  \frac{w}{v} + t^4 )  }{(1 - t^2 v)(1 - t^2/w)(1 - t^2 w/v)(1 - t^3 y)(1 - t^3/y)}.
\]
As noted in \cite{Janik:2007pm},  this  precisely agrees with the $AdS_5 \times S^5$  single particle supergraviton $1/16$th BPS partition function \cite{Gunaydin:1984fk,Kinney:2005ej}, giving a nice confirmation of $AdS$/CFT. Here we see that (\ref{eq:susypartition}) agrees with $\mathcal{I}_{C,\text{ s.t.}}$ up to three sign changes in the denominator and one sign change in the numerator.  These sign changes reflect the charged index's $(-1)^F$ factor and its $(-1)^L$ factor due to $C$.  We conclude that the planar single-trace charged index counts all the the single-trace $1/16$th BPS states graded according to fermion number and $C$ charge, with oscillator parameters replaced by chemical potentials as in (\ref{eq:oscillatortochemicalpotentials}). One can check that the ordinary planar superconformal index follows from the supersymmetric partition function in the analogous way.  However, this is less superficially apparent due to cancellations as the single-trace superconformal index is \cite{Kinney:2005ej}
\< \label{eq:istpolya}
\mathcal{I}_{\text{s.t.}} \eq \sum_{r=1}^{\infty}   \frac{ -\phi(r)}{r} \log(1 - f(x^r))  
\\
\eq   \frac{t^2/w}{1 - t^2/w} + \frac{t^2 v }{1 -  t^2 v} + \frac{t^2 w/v}{1 - t^2 w/v} - 
 \frac{t^3/y}{1 - t^3/y} - \frac{t^3 y}{1 - t^3 y}. \notag
 \>
The first expression for $\mathcal{I}_{\text{s.t.}}$ follows from Polya's theorem. The Euler Phi function $\phi(r)$ gives the number of positive integers less or equal to $r$ that are relatively prime to $r$.

We conclude this section by computing the full planar charged index using the standard formula for a multi-particle partition function (or index) in terms of the single-particle partition function. We have
\< \label{eq:icn=infty}
\log \mathcal{I}_{C, \, N=\infty} \eq \sum_{n \text{ odd}} \frac{\mathcal{I}_{C, \text{ s.t.}}(x^n)}{n} + \sum_{n \text{ even}} \frac{\mathcal{I}_{\text{s.t.}}(x^n)}{n}-  \sum_{n=1}^\infty (-1)^n \frac{ f(x^n)}{n}
\nln
\eq \sum_{n=1}^{\infty} \frac{1}{2}(1 - (-1)^n) \big( \frac{  (f(x^n)-1)^2   }{2 n(1 - f(x^{2 n})) }  - \frac{1}{2 n} \big) 
\nl
+\sum_{n=1}^\infty \sum_{r=1}^{\infty}   \frac{1}{2}(1 + (-1)^n) \frac{ -\phi(r)}{r n} \log(1 - f(x^{r n})) - \sum_{n=1}^{\infty}  (-1)^n f(x^n)/n.
\nl
\>
The first term on the right side of the first line corresponds to odd powers of traces, which retain the sign with respect to $C$ of the single trace. For the second term, which corresponds to even powers of traces, the sign of the single trace with respect to $C$ is irrelevant  since traces repeated even number of times contribute to $\mathcal{I}_C$ in the same way as they would contribute to the ordinary index. Finally, the last term subtracts the single-letter traces to give the charged index for $SU(N)$ instead of $U(N)$. To reach the second equality of (\ref{eq:icn=infty}), we substituted for  $ \mathcal{I}_{C, \text{ s.t.}}$ using (\ref{eq:icst}), and substituted the Polya-method expression for the ordinary single-trace index $\mathcal{I}_{\text{s.t.}}$ (\ref{eq:istpolya}). For the first term on the last line of (\ref{eq:icn=infty}), we substitute $n' = n r /2$ to obtain
\<
\sum_{n=1}^\infty \sum_{r=1}^{\infty}   \frac{1}{2}(1 + (-1)^n) \frac{ -\phi(r)}{r n} \log(1 - f(x^{r n})) \eq -\sum_{n'=1}^\infty  \frac{\sum_{r | n'} \phi(r) (\log (1 - f(x^{2 n'})))}{2 n'}
\nln
\eq -\sum_{n'=1}^\infty \half \log (1 - f(x^{2 n'})),
\>
where the second line follows from the identity
\[
\sum_{b|a} \phi(b)=a,
\]
and the sum is over positive integers $b$ that divide $a$. Now it is straightforward to substitute this simplified term into (\ref{eq:icn=infty}) and exponentiate. We obtain  
\<
\mathcal{I}_{C, \, N=\infty} \eq \prod_{n \text{ odd}} \exp\Big( \frac{  (f(x^n)-1)^2   }{2 n(1 - f(x^{2 n})) }  - \frac{1}{2 n}\Big) 
\nl
\times \prod_{n \text{ even}} \frac{1}{(1 - f(x^n))^{1/2}} \prod_n \exp \Big((-1)^{n+1} \frac{f(x^n)}{n} \Big).
\>
This matches the large-N limit (\ref{eq:largenlimitic}) of the conjectured exact matrix integral expression for $\mathcal{I}_C$ (\ref{eq:matrixintegralforic}).

\section{Conclusions \label{sec:concl}}

We have shown how to compute the charged superconformal index $\mathcal{I}_C$ for $\mathcal{N}=4$ SYM with $SU(N)$ gauge group, and therefore, for general  superconformal theories  with all fields transforming in the adjoint of $SU(N)$. The matrix integral expression (\ref{eq:matrixintegralforic}) for $\mathcal{I}_C$ follows from   (\ref{eq:tracekc=chi}) and (\ref{eq:evaluateddeterminant}), which we have confirmed for small $N$ and for the large-$N$ limit.  Nonetheless, it would be good to prove these formulas. It would be interesting to derive the matrix integral from a path integral approach too, as was done for the partition function in four-dimensions \cite{Aharony:2003sx}, for the index of the three-dimensional superconformal field theory of ABJM \cite{Aharony:2008ug} in \cite{Kim:2009wb}, and for the indices of three-dimensional $\mathcal{N}=2$ superconformal field theories \cite{Imamura:2011su}. Also, we expect that there is a similar story for $\mathcal{I}_C$ for gauge group $E_6$, with the charged character for $E_6$ related to ordinary characters of $F_4$.

We found that $\mathcal{I}_C$ for $\mathcal{N}=4$ SYM is independent of $N$ in the large-$N$ limit. This is not surprising since the ordinary superconformal index is also independent of $N$ in this limit. Nonetheless, this confirms that dynamics need to be taken into account to find the gauge theory duals of $1/16$th BPS black holes in $AdS_5 \times S^5$ \cite{Gutowski:2004yv,Chong:2005da,Kunduri:2006ek}. As suggested by \cite{Berkooz:2008gc}, it is possible that the dual states will only be ``near-BPS,'' with energies suppressed by inverse powers of large charges.

Beyond $\mathcal{N}=4$ SYM, there are many other superconformal theories for which the charged superconformal index can be computed. As one example, $\mathcal{I}_C$ could be computed for Seiberg dual $\mathcal{N}=1$ SQCD theories (with special unitary gauge groups). In this case, the fundamental matter representations would contribute to $\mathcal{I}_C$ only through the adjoint and singlet representations formed by conjugate pairs of representations. Perhaps  $\mathcal{I}_C$ can be related to interesting integral identities such as those of \cite{Spiridonov:2008aa}, as is the case for the ordinary superconformal index. 

The  partition function for three-dimensional superconformal theories on $S^3$ \cite{Kapustin:2009kz,Kapustin:2010xq} has been investigated recently as the dimensional reduction of four-dimensional superconformal indices \cite{Dolan:2011rp,Gadde:2011ia,Imamura:2011uw,Benini:2011nc}. It would be  interesting also to consider the dimensional reduction of the charged superconformal index.

\subsection*{Acknowledgments}
I would like to thank Ofer Aharony, Abhijit Gadde, Christoph Keller, and Vyacheslav Spiridonov for helpful comments.  The research of the author was supported by a Lee A. DuBridge Postdoctoral Fellowship of the California Institute of Technology.  In addition, this work is  supported in part by the DOE grant DE-FG03-92-ER40701.
\appendix

\section{Formulas for $SU(N)$, $SO(2 M+1)$ and $SP(2 M)$ \label{sec:appformulas}}

\subsection{Characters}
 For $SU(N)$, we introduce the $N\times N$ matrix $\alpha^{(N)}(\mathbf{d}; z_k)$ with the $i$-$j$th entry 
\[ \label{eq:definealphamatrix}
\alpha^{(N)}_{ij}(\mathbf{d}; z_k) = z_i^{d'_j+N-j}, \quad d'_j = \sum_{j'=j}^{N-1} d_{j'}.
\]
Note that $\alpha^{(N)}_{iN}(\mathbf{d}; z_k)=1$. Here the $d'_j$ are the partition labels corresponding to the Dynkin labels $\mathbf{d}$.  Then the $SU(N)$ character (\ref{eq:definecharacter}) is
\[ \label{eq:suncharacter}
\chi_{\mathbf{d}}(z_i) = (\prod z_i)^{-\half \sum_{j=1}^{N-1} d_j} \frac{\det \alpha^{(N)}(\mathbf{d}; z_i)}{\det \alpha^{(N)}(\mathbf{0}; z_i)},
\]
where $\mathbf{0}$ has all $(N-1)$ entries equal to $0$. The prefactor is just our convention to ensure that the characters are homogeneous  in the $z_i$, that is $\chi$ is invariant when all $z_i$ are multiplied by a common factor.  This homogeneity is useful for simplifying integration over $SU(N)$. For all the (real) representations relevant for this work, the exponent for the product of the $z_i$ is an integer. In the standard convention, this prefactor is not included and one sets $z_N=(z_1 \ldots z_{N-1})^{-1}$. Note that the powers of $z_i$ in $\chi_{\mathbf{d}}(z_i)$ correspond to the partition labels as  (after setting $z_N=(z_1 \ldots z_{N-1})^{-1}$)
\[
\chi_{\mathbf{d}}(z_i)= \sum_{\mathbf{e}} \prod_{i=1}^{N-1} z_i^{e'_i} .
\]
The sum is over all $\mathbf{e}$ in  $R_{\mathbf{d}}$, and the $e'_i$ and $\mathbf{e}$ are related in the same way as $d'$ and $\mathbf{d}$ are related in (\ref{eq:definealphamatrix}).

Similarly, for $SO(2M+1)$  we introduce the $M\times M$ matrix $\beta^{(M)}(\mathbf{d}; z_k)$ with the $i$-$j$th entry 
\[
\beta^{(M)}_{ij}(\mathbf{d}; z_k) = z_i^{d'_j+M-j+\half}-z_i^{-(d'_j+M-j+\half)}, \quad d'_j = \half d_M + \sum_{j'=j}^{M-1} d_{j'}. \label{eq:definebeta}
\]
Now we have
\[ \label{eq:socharacter}
\chi^{SO(2M+1)}_{\mathbf{d}}(z_i) = \frac{\det \beta^{(M)}(\mathbf{d}; z_i)}{\det \beta^{(M)}(\mathbf{0}; z_i)}.
\]
For all of the (non-spinor) representations relevant for this work the $d'_j$ are integer, and these characters then expand in integer powers of the $z_i$.  This gives a sum
\[
\chi^{SO(2M+1)}_{\mathbf{d}}(z_i) = \sum_{\mathbf{e}} \prod_{i=1}^{M} z_i^{e'_i} ,
\]
where the sum is over all $\mathbf{e}$ in the irreducible $SO(2M+1)$ representation  $R_{\mathbf{d}}$, and the $e'_i$ are related to the $e_i$ as the $d_i'$ and $d_i$ are related in (\ref{eq:definebeta}).

Finally, the analogous expressions for the characters of $SP(2M)$ representation are:  
\begin{gather}
\gamma^{(M)}_{ij}(\mathbf{d}; z_k) = z_i^{d'_j+M-j+1}-z_i^{-(d'_j+M-j+1)}, \quad d'_j = \sum_{j'=j}^{M} d_{j'}, 
\notag \\
\chi^{SP(2M)}_{\mathbf{d}}(z_i) = \frac{\det \gamma^{(M)}(\mathbf{d}; z_i)}{\det \gamma^{(M)}(\mathbf{0}; z_i)}, \quad \chi^{SP(2M)}_{\mathbf{d}}(z_i)  = \sum_{\mathbf{e}} \prod_{i=1}^{M} z_i^{e'_i}. \label{eq:spcharacterformula}
\end{gather}
%

\subsection{Measures}
First we show that integration over $SU(N)$ of any homogeneous function $f(z_i)$ is equivalent to integrating over $U(N)$ with an inverse factor of $ (2 \pi i) \prod  z_i$.   $f(z_i)$  is homogeneous if it is invariant under rescaling of all $N$ $z_i$ by a common phase. As stated before,  this is convenient for the residue calculation of integrals of Section \ref{sec:smallN} and Appendix \ref{sec:appdata}.  We start with
\begin{gather}
\int_{U(N)} \mathrm{d} \mu(z_i) \, \delta\Big(1-\prod z_j\Big) f(z_i) =\frac{1}{2 \pi} \int_0^{2 \pi}\mathrm{d}\phi \, \int_{U(N)} \mathrm{d} \mu(z_i)  \, \delta\Big(1-\prod z_j\Big) f(z_i)
 \notag \\
 = \frac{1}{2 \pi} \int_0^{2 \pi}\mathrm{d}\phi \,e^{-N i \phi}  \, \int_{U(N)} \mathrm{d} \mu(z'_i) \, \delta\Big(1-e^{-N i \phi}\prod z'_j\Big) f(z'_ie^{-i \phi}).
\end{gather}
The first equality is just multiplying and dividing by $2 \pi$, while the second line we changed variables as $z'_i = e^{i \phi} z_i$. Next we integrate out $\phi$ using the delta function, remembering to include the Jacobian factor. Using that $f$ is homogeneous and dropping the primes on variables we obtain
\[
\int_{U(N)} \mathrm{d} \mu(z_i) \, \delta\Big(1-\prod z_j\Big) f(z_i) = \frac{1}{2 \pi i} \int_{U(N)} \frac{\mathrm{d}\mu(z_j)}{\prod z_j}  f(z_i).
\]
Substituting an explicit expression for the Haar measure  of $U(N)$ gives the Haar  measure $\mathrm{d}\mu(z_i)$ for $SU(N)$ for a homogeneous integrand
\[ \label{eq:sunmeasure}
\mathrm{d}\mu(z_i) = \frac{1}{N!} \prod_{j=1}^N \frac{\mathrm{d}z_j}{2 \pi i \, z_j}  \Delta(z) \Delta(z^{-1}).
\]
The $z_i$ are integrated around unit circles and the Vandermonde determinants are
\[
\Delta(z) = \prod_{1 \leq i < j \leq N} (z_i - z_j), \quad \Delta(z^{-1}) = \prod_{1 \leq i < j \leq N} (z_i^{-1} - z_j^{-1}).
\]

For $SO(2M+1)$ we have
\begin{gather}
\mathrm{d}\mu(z_i)= \frac{(-1)^M}{ M!\, 2^M} \prod_{j=1}^M \frac{(z_j + z_j^{-1} - 2 )\, \mathrm{d}z_j}{2 \pi i \, z_j}  \Delta(z + z^{-1})^2 , 
\notag \\
\Delta(z + z^{-1}) = \prod_{i=1}^{M-1}   \prod_{j=i+1}^M (z_i + z_i^{-1}- z_j - z_j^{-1}). \label{eq:someasure} 
\end{gather}
The measure for $SP(2 M)$ is very similar,
\[
\mathrm{d}\mu(z_i)= \frac{(-1)^M}{ M!\, 2^M} \prod_{j=1}^M \frac{(z^2_j + z_j^{-2} - 2 )\, \mathrm{d}z_j}{2 \pi i \, z_j}  \Delta(z + z^{-1})^2. \label{eq:spmeasure} 
\]
The only change is that the  $(z_j+z_j^{-1}-2)$ factor for $SO(2M+1)$ is replaced by $(z_j^2+z_j^{-2}-2)$.

For Section \ref{sec:largeN}, we also need the Haar measures in terms of the angular variables $\theta_j$. Then the measure for $SO(2M+1)$ becomes,
\begin{gather}
\mathrm{d}\mu(\theta_i)= \frac{1}{ M!} \prod_{i=1}^M \frac{\sin^2(\theta_i/2) \, \mathrm{d}\theta_i}{ \pi }  \Delta(\theta)^2 , 
\notag \\
\Delta(\theta) = \prod_{i=1}^{M-1}   \prod_{j=i+1}^M 2(\cos \theta_i - \cos \theta_j),
\end{gather}
and the measure for $SP(2 M)$ is
\[
\mathrm{d}\mu(\theta_i)= \frac{1}{ M!} \prod_{i=1}^M \frac{\sin^2(\theta_i)\, \mathrm{d}\theta_i}{ \pi }  \Delta(\theta)^2 .
\]
%

\section{Direct tests for $SU(2)$ and $SU(3)$ \label{sec:appsu2su3}}
This appendix confirms the relations (\ref{eq:tracekc=chi}) and (\ref{eq:evaluateddeterminant}) by giving explicit matrices  for the adjoint representation of  $SU(2)$ and $SU(3)$. First, for $SU(2)$ the three independent $\gen{J}^A_B$ are
\[
\gen{J}^1_1 = \begin{pmatrix} 0 & 0 & 0 \\ 0 & -1 & 0 \\  0 & 0 & 1\end{pmatrix}, \quad  \gen{J}^1_2 = \begin{pmatrix} 0 & 0 & -\sqrt 2 \\ \sqrt 2 & 0 & 0 \\ 0 & 0 & 0\end{pmatrix},\quad \gen{J}^2_1 =  \begin{pmatrix} 0 & \sqrt 2 & 0 \\ 0 & 0 & 0 \\  -\sqrt 2 & 0 & 0\end{pmatrix},
\]
and the charge conjugation operator acts in the adjoint representation as
\[
C=  \begin{pmatrix} -1 & 0 & 0 \\ 0 & 0 & -1 \\  0 & -1 & 0\end{pmatrix}.
\]
For this case, $K_C$ is a function of a single variable,
\[
K_C(z_1) = \exp \Big ((z_1^{1/2}+z_1^{-1/2}) \gen{J}^1_2 \Big) C.
\]
Direct evaluation than gives
\[
\mathrm{Tr}K_c(z_1) =  z_1+z_1^{-1}+1, \quad \mathrm{det}(1 - x K_c) = (1- x)(1 - x z_1)(1- \frac{x}{z_1}),
\]
in agreement with (\ref{eq:tracekc=chi}) and (\ref{eq:evaluateddeterminant}) for $N=2$.

Similarly, for $SU(3)$, four of the $\gen{J}^A_B$ are
\begin{align}
\gen{J}^1_1 &= \begin{pmatrix} 
0 & 0 & 0 & 0 & 0 & 0 & 0 & 0 \\  
0 & -1 & 0 & 0 & 0 & 0 & 0 & 0 \\
0 & 0 & -1 & 0 & 0 & 0 & 0 & 0 \\
0 & 0 & 0 & 1 & 0 & 0 & 0 & 0 \\
0 & 0 & 0 & 0 & 0 & 0 & 0 & 0 \\
0 & 0 & 0 & 0 & 0 & 0 & 0 & 0 \\
0 & 0 & 0 & 0 & 0 & 0 & 1 & 0 \\
0 & 0 & 0 & 0 & 0 & 0 & 0 & 0 
\end{pmatrix}, &
\gen{J}^1_2 &= \begin{pmatrix} 
0 & 0 & 0 & -\frac{1}{\sqrt 2} & 0 & 0 & 0 & 0 \\  
\frac{1}{\sqrt 2} & 0 & 0 & 0 & -\sqrt{\frac{3}{2}} & 0 & 0 & 0 \\
0 & 0 & 0 & 0 & 0 & -1 & 0 & 0 \\
0 & 0 & 0 & 0 & 0 & 0 & 0 & 0 \\
0 & 0 & 0 & \sqrt{\frac{3}{2}} & 0 & 0 & 0 & 0 \\
0 & 0 & 0 & 0 & 0 & 0 & 0 & 0 \\
0 & 0 & 0 & 0 & 0 & 0 & 0 & 0 \\
0 & 0 & 0 & 0 & 0 & 0 & 1 & 0 
\end{pmatrix}, \notag \\
\gen{J}^2_2 &= \begin{pmatrix} 
0 & 0 & 0 & 0 & 0 & 0 & 0 & 0 \\  
0 & 1 & 0 & 0 & 0 & 0 & 0 & 0 \\
0 & 0 & 0 & 0 & 0 & 0 & 0 & 0 \\
0 & 0 & 0 & -1 & 0 & 0 & 0 & 0 \\
0 & 0 & 0 & 0 & 0 & 0 & 0 & 0 \\
0 & 0 & 0 & 0 & 0 & -1 & 0 & 0 \\
0 & 0 & 0 & 0 & 0 & 0 & 0 & 0 \\
0 & 0 & 0 & 0 & 0 & 0 & 0 & 1 
\end{pmatrix}\, &
\gen{J}^2_3 &= \begin{pmatrix} 
0 & 0 & 0 & 0 & 0 & 0 & 0 &  -\frac{1}{\sqrt 2} \\  
0 & 0 & 0 & 0 & 0 & 0 & 0 & 0 \\
0 & 1 & 0 & 0 & 0 & 0 & 0 & 0 \\
0 & 0 & 0 & 0 & 0 & 0 & -1 & 0 \\
0 & 0 & 0 & 0 & 0 & 0 & 0 & -\sqrt{\frac{3}{2}} \\
\frac{1}{\sqrt 2} & 0 & 0 & 0 & \sqrt{\frac{3}{2}} & 0 & 0 & 0 \\
0 & 0 & 0 & 0 & 0 & 0 & 0 & 0 \\
0 & 0 & 0 & 0 & 0 & 0 & 0 & 0 
\end{pmatrix}.
\end{align}
We can build the other four generators in terms of these four. We have $\gen{J}^1_3 = \comm{\gen{J}^2_3}{\gen{J}^1_2}$, and then the remaining generators are given by Hermitian conjugation (they are the transpose matrices),
\[
\gen{J}^2_1 = (\gen{J}^1_2)^\dagger, \quad \gen{J}^3_2 = (\gen{J}^2_3)^\dagger, \quad \gen{J}^3_1 = (\gen{J}^1_3)^\dagger.
\]
Now the charge conjugation operator is represented by
\[
C = \begin{pmatrix} 
-1 & 0 & 0 & 0 & 0 & 0 & 0 & 0 \\  
0 & 0 & 0 & -1 & 0 & 0 & 0 & 0 \\
0 & 0 & 0 & 0 & 0 & 0 & -1 & 0 \\
0 & -1 & 0 & 0 & 0 & 0 & 0 & 0 \\
0 & 0 & 0 & 0 & -1 & 0 & 0 & 0 \\
0 & 0 & 0 & 0 & 0 & 0 & 0 & -1 \\
0 & 0 & -1 & 0 & 0 & 0 & 0 & 0 \\
0 & 0 & 0 & 0 & 0 & -1 & 0 & 0 
\end{pmatrix}.
\]
Again, $K_C$ is built from a single raising operator,
\[
K_C(z_1) = \exp \Big ((z_1^{1/2}+z_1^{-1/2}) \gen{J}^1_2 \Big) C,
\]
and again we find agreement with  (\ref{eq:tracekc=chi}) and (\ref{eq:evaluateddeterminant}), this time for $N=3$,
\[
\mathrm{Tr}K_c(z_1) =  z_1+z_1^{-1}, \quad \mathrm{det}(1 - x K_c) = (1- x^2)(1 - x z_1)(1- \frac{x}{z_1})(1 + x^2 z_1)(1+ \frac{x^2}{z_1}).
\]
%
\section{Further tests for small $N$\label{sec:appdata}}

In this section we report the computations of $I^\pm$ and $I^\pm_C$ for $N=4,5$, which provide additional confirmation of the charged character relations (\ref{eq:tracekc=chi}) and the determinant expression (\ref{eq:evaluateddeterminant}).

 For $SU(4)$ we have
\< 
\Phi(a_1,a_2;z_1,z_2,z_3,z_4) & \rightarrow & \, 4! \frac{z_1^3 (z_2 z_3 z_4)^2}{\Delta(z)(a_1 z_1 - z_4)(a_2^2 z_1 z_2 - z_3 z_4)}, 
\nln
I^-(x,a_1, a_2) \eq \frac{\big((1 - x^2)(1 - x^3)(1 - x^4)\big)^{-1}(1 + a_1 a_2^2 x^6)}{(1 - a_1 x)(1 - a_1 x^2)(1 - a_1 x^3)(1 - a_2^2 x^2)(1 - a_2^2 x^4)},
\notag \\
I^+(x,a_1, a_2)\eq (1 - x)^3 \Big((1 + x^{12}) + (x + x^{11}) (3 - a_1 ) + ( 
      x^2 + x^{10}) (6 - 2 a_1 ) 
      \nl
      + (x^3 + x^9) (9 - 4 a_1 - a_1^2 - a_2^2 - a_1 a_2^2 ) 
   \nl
    + (x^4 + x^8) (12 - 5 a_1 - a_1^2 - a_2^2 - a_1 a_2^2 + 
       a_2^4) 
          \nl + (x^5 + x^7) (14 - 7 a_1 - 2 a_1^2 - a_1^3 - a_2^2 - 
       3 a_1 a_2^2 + 2 a_2^4 ) 
       \nl
       +  x^6 (15 - 7 a_1 - a_1^2 - a_1^3 - 2 a_1 a_2^2 + a_1^2 a_2^2 + 
       3 a_2^4)     \Big) .
\>
Comparing to the results for $G(4) = SO(5)$, 
\< 
\lefteqn{\Theta^{SO(5)}(a_1,a_2;z_1,z_2)  \rightarrow -2 (2!) \frac{(z_1 z_2)^2}{(z_1-z_2)(1-z_2)(1 - z_1 z_2)} } 
\notag \\
\lefteqn{\times \frac{(-1-z_1 + a_1 z_1 -z_1^2 + a_2^2 z_1 z_2 + a_1 a_2^2 z_1 z_2)}{(a_1-z_1)(1-a_1 z_1)(z_1-a_2^2 z_2)(1 - a_2^2 z_1 z_2)},}
\nln
I_C^-(x,a_1, a_2) \eq \frac{\big((1 - x^2)(1 + x^3)(1 - x^4)\big)^{-1}(1 + a_1 a_2^2 x^6)}{(1 - a_1 x)(1 + a_1 x^2)(1 - a_1 x^3)(1 - a_2^2 x^2)(1 - a_2^2 x^4)},
\notag \\
I_C^+(x,a_1, a_2)\eq (1-x)(1-x^2)\Big(1+ x^{12} + (x+x^{11})(1 - a_1) + 2 (x^2 + x^{10}) 
 \nl
+ (x^3 +x^9)(1 - a_1^2  - a_2^2  + a_1 a_2^2)    +  x^6(3 - a_1  + a_1^2  - a_1^3  + a_1^2 a_2^2  - a_2^4)
\nl
+  (x^4+x^8)(2 + a_1  - a_1^2  -a_2^2  + a_1 a_2^2  - a_2^4) 
\nl
 +  (x^5 +x^7)(2 - a_1  - a_1^3  - a_2^2 + a_1 a_2^2)\Big),
\>
again we find consistency with (\ref{eq:nmcondition}).

Finally, for $N=5$ we report the results for $I^\pm$,
\<
I^-(x, a_1, a_2) \eq \frac{(1- a_1^2 a_2 x^{10})(1 +  a_2^2 x^{10}) + (1 - a_2 x^5) (a_1 a_2(x^6 + x^7 + x^8 + x^9))   }{(1 - a_1 x)(1 - a_1 x^2)(1 - a_1 x^3)(1 - a_1 x^4)}
 \nl
 \times \frac{\big((1 - x^2)(1 - x^3)(1 - x^4)(1 - x^5)\big)^{-1}}{(1 - a_2 x^2)(1 - a_2 x^3)(1 - a_2 x^4)(1 - a_2 x^5)(1 - a_2 x^6)},
\notag \\
I^+(x, a_1, a_2)  \eq (1 - x)^4 \Big((1 + 
      x^{20}) + (x + x^{19}) (4 - a_1 ) + (x^2 +  x^{18}) (10 - 
       3 a_1 )
       \nl
        + (x^3 +  x^{17}) (19 - 7 a_1 - a_1^2 - a_2 - 
       a_1 a_2 )
       \nl
        + (x^4 +  x^{16}) (31 - 12 a_1 - 2 a_1^2 - 2 a_2 - 2 a_1 a_2 +
        a_2^2 )
        \nl
         + (x^5 +  x^{15}) (45 - 19 a_1 - 4 a_1^2 - a_1^3 - 4 a_2 - 
       6 a_1 a_2 - a_1^2 a_2 + 3 a_2^2) 
       \nl
       + (x^6 +  x^{14}) (60 - 26 a_1 - 5 a_1^2 - 2 a_1^3 - 5 a_2 - 8 a_1 a_2 - a_1^2 a_2 + 7 a_2^2 +     a_1 a_2^2 ) 
       \nl
       + (x^7 +  x^{13}) (74 - 34 a_1 - 8 a_1^2 - 4 a_1^3 - 
       a_1^4 - 7 a_2 
       \nl
       - 14 a_1 a_2 - 4 a_1^2 a_2 + 9 a_2^2 + 
       a_1 a_2^2)
       \nl
        + (x^8 +  x^{12}) (86 - 40 a_1 - 9 a_1^2 - 4 a_1^3 - 
       2 a_1^4 - 7 a_2 - 14 a_1 a_2 - 3 a_1^2 a_2  
       \nl 
      + a_1^3 a_2 + 13 a_2^2 + 4 a_1 a_2^2 + a_2^3) 
       \nl
       + (x^9 +  x^{11}) (94 - 45 a_1 - 11 a_1^2 - 
       5 a_1^3 - 3 a_1^4 - 8 a_2 - 19 a_1 a_2 - 7 a_1^2 a_2 
       \nl
    + a_1^3 a_2  +  14 a_2^2    + 3 a_1 a_2^2 + 2 a_2^3) 
       \nl
        +  x^{10} (97 - 46 a_1 - 10 a_1^2 - 4 a_1^3 - 3 a_1^4 - 7 a_2 - 16 a_1 a_2 - 
       4 a_1^2 a_2 + 2 a_1^3 a_2 
       \nl
     + 17 a_2^2   + 6 a_1 a_2^2 + a_1^2 a_2^2 + 
       3 a_2^3) \Big) ,
\>
and for $I^\pm_C$,
\<
I^-_C(x, a_1, a_2) \eq \frac{1 - a_2 x^5 + a_1 a_2 x^6 - a_1 a_2 x^7 + a_1 a_2 x^8 - a_1 a_2 x^9 + 
 a_1^2 a_2 x^{10} - a_1^2 a_2^2 x^{15}}{(1 - a_1 x)(1 + a_1 x^2)(1 - a_1 x^3)(1 + a_1 x^4)}
 \nln
& & \times \frac{\big((1 - x^2)(1 + x^3)(1 - x^4)(1 + x^5)\big)^{-1}}{(1 - a_2 x^2)(1 +  a_2 x^3)(1 - a_2 x^4)(1 - a_2 x^6)},
\notag \\
I^+_C(x, a_1, a_2)  \eq (1-x^2)^2\Big(1+ x^{20} - a_1 (x+x^{19}) + (x^2 + x^{18})  (2+a_1)
 \nl
- (x^3 +x^{17})(1 + a_1 + a_1^2 + a_2  -  a_1 a_2)  +  (x^4+x^{16})(3 + 2 a_1  - a_2^2)
 \nl
  -  (x^5 +x^{15})(1 + 3 a_1  + a_1^3  -a_1^2 a_2 - a_2^2 ) 
  \nl
  +  (x^6+ x^{14})(4 + 2 a_1  + a_1^2  + a_2 - 2 a_1 a_2 + a_1^2 a_2 - a_2^2 - a_1 a_2^2 )
  \nl
  - (x_7 + x^{13})(2 + 2 a_1 + 2 a_1^2 + a_1^4 + a_2 - 2 a_1 a_2  - a_2^2 - a_1 a_2^2) 
  \nl
  + (x^8 + x^{12})(4 + 4 a_1 -a_1^2 + 2 a_1^3 - a_2 -a_1^2 a_2 + a_1^3 a_2 - 3 a_2^2 - a_2^3)
  \nl
  -(x_9 + x^{11})(2 + 3 a_1 + a_1^2 + a_1^3 + a_1^4 - a_1 a_2 - a_1^2 a_2 + a_1^3 a_2 - 2 a_2^2 - a_1 a_2^2) 
  \nl
  + x^{10}(5 + 2 a_1 + 2 a_1^2 + a_1^4 + a_2 - 4 a_1 a_2 + 2 a_1^2 a_2 - a_2^2
  \nl
   - 2 a_1 a_2^2 + a_1^2 a_2^2 - a_2^3)\Big). 
\>
It is straightforward to check that the fermionic $I^+$ and $I^+_C$ satisfy (\ref{eq:nmcondition}). However, the check for  $I^-$ and $I^-_C$ requires more work. 
First note that $I^-_C$  expands with all positive coefficients if we simply flip the signs of $x$ and $a_1$,
\<
I^-_C(-x, -a_1, a_2) \eq \frac{\prod_{i=2}^5 (1-x^i)^{-1}}{(1 - a_1 x)(1 -  a_2 x^3)(1 - a_2 x^4)}  \bigg(\frac{1}{(1 - a_1 x^2)(1 - a_2 x^2)(1 - a_2 x^6) } 
\nl
  + \frac{ x^3 a_1}{(1 - a_1 x^2)(1 - a_1 x^3)(1 - a_2 x^2) }  + \frac{ x^4 a_1 }{(1 - a_1 x^2)(1 - a_1 x^4)(1 - a_2 x^6) }
\nl
 +\frac{x^7 a_1^2}{(1 - a_1 x^2)(1 - a_1 x^3)(1 - a_1 x^4) }
 \nl
 + \frac{x^5 a_2 }{(1 - a_1 x^4)(1 - a_2 x^2)(1 - a_2 x^6) } \bigg). 
\>
Subtracting this from $I^-$ and simplifying leads to
\<
\lefteqn{I^-(x, a_1, a_2) - I^-_C(-x, -a_1, a_2)  = 2  x^6 \frac{\prod_{i=2}^5 (1-x^i)^{-1}}{(1 - a_1 x^2)(1 - a_1 x^3)}  }
\notag \\
& & \times   \frac{a_2}{(1 -  a_2 x^2)(1 -  a_2 x^3)(1 - a_2 x^4)(1 -  a_2 x^6)}  
\notag \\
& & \times \bigg(\frac{a_1}{(1 - a_1 x)(1 - a_2 x^5) } + \frac{a_2 x^4}{(1 - a_1 x^4)(1 - a_2 x^5) } + \frac{a_1(x+x^2+x^3)}{(1 - a_1 x)(1 - a_1 x^4) } \bigg),
\>
which clearly expands with all even positive integer coefficients, as required.


\end{document}